\documentclass[aps,twocolumn,superscriptaddress,floatfix]{revtex4-1}

\usepackage{booktabs}
\usepackage{lipsum}
\usepackage{graphicx}
\usepackage{amsmath,amssymb}
\usepackage{braket}
\usepackage{mathtools}
\usepackage{hyperref}
\usepackage{color}
\usepackage[normalem]{ulem}

\newcommand{\Z}{\mathbb{Z}}
\newcommand{\C}{\mathbb{C}}
\newcommand{\n}[1]{\left| #1 \right|}
\newcommand{\st}[1]{\left\{ #1 \right\}}

\renewcommand{\v}[1]{\boldsymbol{#1}}
\DeclareMathOperator{\Tr}{Tr}

\usepackage{tikz}
\usetikzlibrary{shapes,calc}
\usetikzlibrary{shapes.geometric,arrows.meta,decorations.markings}

\tikzset{
	dot/.style={draw,circle,inner sep=1.5pt,fill=black},
	empty dot/.style={draw,circle,inner sep=1.5pt,fill=white},
	mid arrow/.style={postaction={decorate,decoration={
        markings,
        mark=at position .6 with {\arrow[#1,scale=1.5]{latex}}
    }}},
	spinA/.style={draw=black,thick,circle,inner sep=2.5pt, fill=figBlue},
	spinBC/.style={draw=black,thick,circle,inner sep=2.5pt, fill=figRed},
	faded/.style={opacity=0.2},
}

\newcommand{\ibz}{\int [d\v{k}]}
\newcommand{\re}{\operatorname{Re}}

\begin{document}

\title{Diagrammatic approach to nonlinear optical response with application to Weyl semimetals}

\author{Daniel E. Parker}
\email[]{daniel\_parker@berkeley.edu}
\affiliation{Department of Physics, University of California, Berkeley, CA 94720, USA}
\author{Takahiro Morimoto}
\email[]{tmorimoto@berkeley.edu}
\affiliation{Department of Physics, University of California, Berkeley, CA 94720, USA}
\affiliation{Materials Science Division, Lawrence Berkeley National Laboratory, Berkeley, California 94720, USA}
\author{Joseph Orenstein}
\email[]{jworenstein@lbl.gov}
\affiliation{Department of Physics, University of California, Berkeley, CA 94720, USA}
\affiliation{Materials Science Division, Lawrence Berkeley National Laboratory, Berkeley, California 94720, USA}
\author{Joel E. Moore}
\email[]{jemoore@berkeley.edu}
\affiliation{Department of Physics, University of California, Berkeley, CA 94720, USA}
\affiliation{Materials Science Division, Lawrence Berkeley National Laboratory, Berkeley, California 94720, USA}

\newcommand{\dtau}{d\v{\tau}^{(3)}}

\date{\today}

\begin{abstract}
Nonlinear optical responses are a crucial probe of physical systems including periodic solids.  In the absence of electron-electron interactions, they are calculable with standard perturbation theory starting from the band structure of Bloch electrons, but the resulting formulas are often large and unwieldy, involving many energy denominators from intermediate states. 
This work gives a Feynman diagram approach to calculating non-linear responses. This diagrammatic method is a systematic way to perform perturbation theory, which often offers shorter derivations and also provides a natural interpretation of nonlinear responses in terms of physical processes.  Applying this method to second-order responses concisely reproduces formulas for the second-harmonic, shift current.
We then apply this method to third-order responses and derive formulas for third-harmonic generation and self-focusing of light, which can be directly applied to tight-binding models.  Third-order responses in the semiclasscial regime include a Berry curvature quadrupole term, whose importance is discussed including symmetry considerations and when the Berry curvature quadrupole becomes the leading contribution.  The method is applied to compute third-order optical responses for a model Weyl semimetal, where we find a new topological contribution that diverges in a clean material, as well as resonances with a peculiar linear character.
\end{abstract}
\maketitle

\tableofcontents

\section{Introduction}

Optical response provides a window into the quantum nature of materials.  The exquisite control and precise measurements enabled by modern optical techniques frequently couple with theoretical predictions to test and confirm models of quantum materials.  Nonlinear optical responses \cite{Boyd,Bloembergen,butcher1991elements,Sturman}, in particular, give a wealth of information on dynamics, symmetry, and---recently---topology \cite{sipe2000second,Young-Rappe,Young-Zheng-Rappe,Cook17,Morimoto-Nagaosa16,orensteinmoore,sodemannfu,morimoto2016semiclassical,Chan16,de2017quantized,Wu17,ma2017direct,Patankar}. To fully reap the benefits of optical techniques, it is necessary to accurately predict optical responses in solids from theory, including both simplified tight-binding models and advanced computational approaches.

Historically, optical responses were understood first at the linear order, and then extended to nonlinear orders alongside the development of the laser in the 1960s. For molecular systems, normal quantum-mechanical perturbation theory suffices, and a convenient diagrammatic language became popular, capturing optical processes in terms of electrons changing energy levels \cite{Ward65}. In crystalline systems, however, there are several additional wrinkles. Simply defining the perturbation corresponding to an external electric field is a subtle task. Like in a molecule, absorbing a photon can cause an electron to jump to a different band, but can also cause the electron to move to a nearby $k$-point on the same band. The latter requires connecting adjacent points in $k$-space, and thus involves the Berry connection \cite{berry1984quantal,simon1983holonomy}.

It is only relatively recently that the electromagnetic perturbation was written carefully in order to treat nonlinear responses. There are two standard ways of writing an electromagnetic perturbation within the framework of independent electrons and dipole fields. First, the so-called \textit{length gauge}
\begin{equation}
	\widehat{H}_E = \widehat{H}_0 + e \v{E}(t) \cdot \widehat{\v{r}}
	\label{eq:length_gauge}
\end{equation}
uses the single-particle position operator $\widehat{\v{r}}$ whereas the second, the \textit{velocity gauge}, uses the minimal substitution scheme
\begin{equation}
	\widehat{H}_A = \widehat{H}_0(\v{k}-e\v{A}(t))
	\label{eq:velocity_gauge}
\end{equation}
where the vector potential $\v{A}(t)$ is chosen so that $\v{E}(t) = - \partial_t \v{A}(t)$. As usual, each gauge is well-suited for a different set of tasks. The length gauge is better for analytical answers, semi-classical limits, and some questions involving topology, whereas the velocity gauge gives a cleaner resonance structure and is easier to implement numerically, especially for tight-binding models.

Over time, there has been a competition between the two approaches. The velocity gauge was initially favored in the 1980s due to easier calculation (see e.g. \cite{kraut1979anomalous,von1981theory}). Velocity gauge calculations, however, often contain spurious divergences at zero frequency, which can be eliminated only with somewhat opaque sum rules. The position operator was defined carefully in the work of Blount in the 1960s \cite{blount1962formalisms}, and its relation to the Berry connection was understood deeply by the early 1990s, giving rise to the modern theory of polarization \cite{Resta,Kingsmith93}. This understanding was harnessed by Sipe and Shkrebtii to develop a widely used approach to calculate second-order responses within the length gauge \cite{sipe2000second}. The wide variety of physical effects in the second-order response---including shift current \cite{Young-Rappe,Young-Zheng-Rappe,Cook17,Morimoto-Nagaosa16}, injection current \cite{Chan16,Taguchi16,de2017quantized}, and second-harmonic generation (SHG) \cite{Morimoto-Nagaosa16,Yang17}---are of great current interest for a wide variety of systems. These convenient formulae \cite{sipe2000second}, together with putative dangers associated with the velocity gauge when truncating the number of bands, ensured the primacy of the length gauge.

Recent work \cite{ventura2017gauge,passos2017nonlinear,taghizadeh2018gauge} has re-examined the roots of the problem, providing careful prescriptions for both gauges and how to translate between them. It is now possible to use either gauge correctly, depending on the problem at hand. In this work we focus primarily on the relatively underappreciated velocity gauge, developing a convenient Feynman diagram prescription for calculating nonlinear responses.  As noted above, diagrammatic methods have a long history in the subject. The formulation here, however, has several key differences from previous work to implement the correct form of the electromagnetic interaction and fully account for the effects of the Berry connection. Our goal is to show that any second- or third-order optical response can be calculated from diagrams in only a few lines. 
Two practical advantages of the resulting velocity-gauge expressions is that the resonance structure is manifest, immediately distinguishing one-, two-, and three-photon terms, and the expressions can be directly implemented in tight-binding models without the need for sum rules.

One motivation for this work is providing tools to better understand optical responses. A large body of recent work has followed the theme of studying optical responses in situations where they become particularly simple: semiclassics and Weyl semimetals. In semiclassics, the limit of a single band where $\omega \to 0$, optical responses can be understood from the semiclassical equations of motion (EOM) that describe wavepackets of Bloch electrons. Berry curvature gives an additional contribution to these equations of motion called the anomalous velocity, which leads to the Hall conductivity in linear response \cite{nagaosaahereview}. At second order, the anomalous velocity is responsible for the circular photogalvanic effect (CPGE), and non-linear Kerr rotation that is proportional to the dipole of Berry curvature \cite{orensteinmoore,sodemannfu,morimoto2016semiclassical}. 

The next-to-simplest situation is that of two-band models, where interband responses give rise to resonances. Perhaps the most intriguing two-band models are those for Weyl semimetals. These materials support three-dimensional gapless points called Weyl nodes that are sources and sinks of Berry curvature \cite{Murakami,Wan11}. Simple tight-binding models often suffice to describe their properties. However, the fact that the Fermi surface vanishes at the Weyl nodes puts them firmly beyond the semiclassical regime. Due to their nontrivial Berry curvature structure, Weyl semimetals host a variety of non-trivial linear responses, including the chiral magnetic effect \cite{Fukushima,Son} and gyrotropic magnetic effect\cite{zhong2016gyrotropic,ma2015chiral}. As one might expect, there is an even richer set of nonlinear optical responses due to the Berry curvature\cite{de2017quantized,Chan16,Yang17}. These effects can be realized, for instance, in the monopnictide TaAs, a Weyl semimetal with inversion breaking \cite{Weng15,Huang15}. Recent optical experiments on TaAs revealed that TaAs shows CPGE responses closely tied to its Weyl node structure \cite{ma2017direct} and giant SHG, with the largest $\chi^{(2)}$ of any known material\cite{Wu17,Patankar}.

Below we connect the Feynman diagram formulation of optical response to both semiclassics and Weyl semimetals. In the semiclassical limit we show that, with particular symmetries, the largest term in the third-order response has a topological origin as the quadrupole of the Berry Curvature. We also examine the third-harmonic response of a Weyl semimetal. We find that the off-diagonal component $\sigma^{zxxx}$ has large two-photon and three-photon resonances with peculiar linear profiles due to the Weyl cones.

The remainder of this paper is organized as follows. Section \ref{sec:setup} introduces notation and the Feynman rules. Sections \ref{sec:first_order}-\ref{sec:third_order} derive non-linear optical responses through third order using Feynman diagrams and provide some physically interesting limits. Section \ref{sec:semiclassics} considers the semiclassical limit, its relation to the length gauge, and some topological considerations at third order. Section \ref{sec:numerical_example} presents a numerical example of a Weyl semimetal and, lastly, Section \ref{sec:conclusions} concludes with some heuristic rules for choosing a gauge, and other comments.

\section{Setup \& Feynman Rules}
\label{sec:setup}

\begin{table}
\begin{ruledtabular}
\begin{tabular}{ccc}
	\toprule
	Component & Diagram & Value\\[2em]
	\hline\\
\begin{minipage}[h]{2.2cm}
	(Classical) \\
	Photon\\
	Propagator
\end{minipage}&
\begin{tikzpicture}[baseline=(a.center)]
	\node (a) {\includegraphics{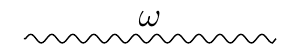}};
\end{tikzpicture}
&
1\\[2em]
\begin{minipage}[h]{2.2cm}
	Electron\\
	Propagator
\end{minipage}&
\begin{tikzpicture}[baseline=(a.center)]
	\node (a) {\includegraphics{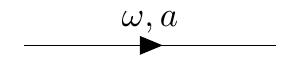}};
\end{tikzpicture}
&
$\displaystyle G_a(\omega)$\\[2em]
\begin{minipage}[h]{2.2cm}
	One-Photon\\
	Input Vertex
\end{minipage}&
\begin{tikzpicture}[baseline=(a.center)]
	\node (a) {\includegraphics{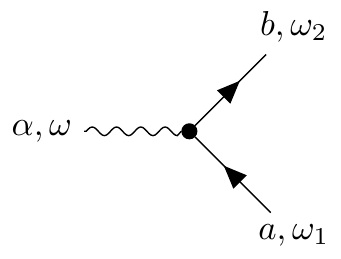}};
\end{tikzpicture}
&
$\frac{ie}{\hbar \omega_1} h^\alpha_{ab}$\\[2em]
\begin{minipage}[h]{2.2cm}
	Two-Photon\\
	Input Vertex
\end{minipage}&
\begin{tikzpicture}[baseline=(a.center)]
	\node (a) {\includegraphics{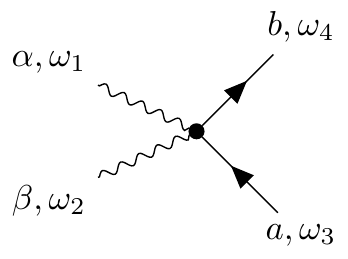}};
\end{tikzpicture}
&
${\displaystyle \prod_{k=1}^2} \left( \frac{ie}{k \hbar \omega_k} \right)  h^{\alpha\beta}_{ab}$\\[2em]
\begin{minipage}[h]{2.2cm}
	Three-Photon\\
	Input Vertex
\end{minipage}&
\begin{tikzpicture}[baseline=(a.center)]
	\node (a) {\includegraphics{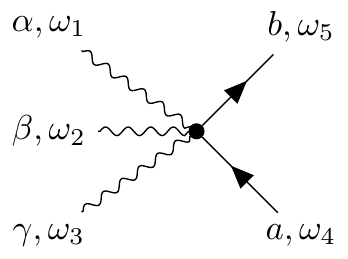}};
\end{tikzpicture}
&
${\displaystyle \prod_{k=1}^3} \left( \frac{ie}{k \hbar \omega_k} \right) h^{\alpha\beta\gamma}_{ab}$\\[2em]
\begin{minipage}[h]{2.2cm}
	One-Photon\\
	Output Vertex
\end{minipage}&
\begin{tikzpicture}[baseline=(a.center)]
	\node (a) {\includegraphics{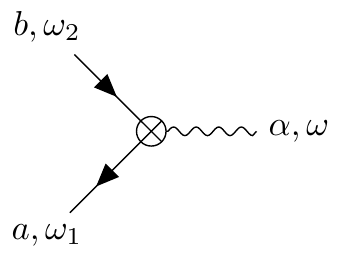}};
\end{tikzpicture}
&
$e h^\mu_{ab}$\\
\end{tabular}
\end{ruledtabular}
\caption{The Feynman rules for non-linear electromagnetic perturbations in a crystal. Following the pattern, a new vertex with $N$ incoming photons will appear at $N$\textit{th} order. Energy must be integrated around each internal loop, and conserved at each vertex. The output vertex can appear with any number of photons and gains a power of $ie (k \hbar\omega_k)^{-1}$ for each additional external photon.}
\label{tab:feynman_rules}
\end{table}

We will work in a band theory picture of non-interacting electrons for simplicity, though most of the considerations involved carry over to Fermi liquid theory. We first recall some key definitions to set notation, then discuss perturbation theory in an external electric field for velocity gauge, and derive the Feynman rules.  We also comment on the assumptions and caveats of the framework.s

\subsection{Band Theory and Notation}

Consider a crystalline material in $d$ dimensions described by band theory. The second-quantized Hamiltonian is then
\begin{equation}
	\widehat{H}_0	= \sum_{a \in \Z} \int [d\v{k}]\;  \varepsilon_a(\v{k}) c^\dagger_{\v{k}a} c_{\v{k}a}
	\label{eq:band_structure}
\end{equation}
where $\int [d\v{k}] = (2\pi)^{-d} \int  d^d \v{k}$ indicates the properly-normalized integral over the $d$-dimensional first Brillouin zone, the sum runs over all bands, and the $c^\dagger$ and $c$'s are single-particle fermion creation and annihilation operators, satisfying the usual anticommutation relations
\begin{equation}
	\{c_{\v{k}a}, c^\dagger_{\v{k'}b}\} = (2\pi)^d \delta_{ab} \delta(\v{k}-\v{k}').
\end{equation}
(Latin indices $a,b,c,d$ are used to label bands henceforth.) We assume that the crystal is infinite in extent, without boundary.

Because the Hamiltonian \eqref{eq:band_structure} involves only a single raising and lowering operator, Fermion number is a symmetry of the system. We may thus write all observables in terms of the single-particle wavefunctions. As usual, Bloch's Theorem says the single-electron wavefunctions may be written as
\begin{equation}
	\psi_{\v{k}a}(\v{r}) = \braket{0|\widehat{\Psi}(\v{r}) c^\dagger_{\v{k}a}|0} = e^{i \v{k}\cdot \v{r}} u_{\v{k}a}(\v{r}),\quad a \in \Z, k \in \text{BZ}
	\label{eq:single_particle_wavefunctions}
\end{equation}
where $\widehat{\Psi}(\v{r})$ annihilates an electron at $\v{r}$, and the $u$'s are periodic functions on the unit cell~\footnote{We note that the Bloch wavefunctions are not necessarily periodic in $\v{k}$. In fact, the natural gauge choice is given by the convention $u_{\v{k}+\v{G}a}(\v{r}) = e^{-i\v{G}\cdot \v{r}} u_{\v{k}a}(\v{r})$. This gauge choice should be adopted when trying to compute polarization and related quantities in tight-binding models.}. 
The $u$'s are eigenfunctions of the $k$-dependent Hamiltonian $\widehat{H}_0(\v{k}) = e^{-i\v{k}\cdot\v{r}} \widehat{H}_0 e^{i\v{k}\cdot \v{r}}$:
\begin{equation}
	\widehat{H}_0(\v{k}) \; u_{\v{k} a}(\v{r}) = \varepsilon_{\v{k}_a} u_{\v{k}a}(\v{r}).
	\label{eq:single_particle_eigenvalues}
\end{equation}
Despite our assumption of independent electrons, we work in a fully many-body framework. This is necessary to implement Feynman diagrams but also, as discussed below, makes the generalization to the interacting case transparent.

\subsection{Electromagnetic Interactions}

Suppose there is an external electric field which we treat classically. We adopt the velocity gauge with the minimal coupling Hamiltonian \eqref{eq:velocity_gauge}. To capture nonlinear responses, we expand in powers of the vector potential in a Taylor series. This is, however, not as straightforward as it might naively seem because one must carefully consider what notion of derivative should be employed in the series. The answer is that one should use the (Berry) covariant derivative $\widehat{D}$ when working in $k$-space. The derivation of this fact and the defintion of the covariant derivative are reviewed in Appendix \ref{app:connections}. Appendix \ref{app:connections} shows that the covariant derivative $\widehat{\v{D}}$ is related to the single-electron position operator via $\widehat{\v{r}} = i \widehat{\v{D}}$, and that it acts naturally on operators via
\begin{equation}
	\widehat{\v{D}}[ \widehat{\mathcal{O}} ] = [ \widehat{\v{D}}, \widehat{\mathcal{O}} ],
	\label{eq:convariant_derivative_on_operators}
\end{equation}
where the commutator has matrix elements
\begin{equation}
	[ \widehat{D}^\mu, \widehat{\mathcal{O}} ]_{ab} = \frac{\partial \mathcal{O}_{ab}}{\partial k^\mu} - i [ \mathcal{A}^\alpha, \widehat{\mathcal{O}} ]_{ab}
	\label{eq:operator_derivative_matrix_elements}
\end{equation}
where $\v{\mathcal{A}}$ is the Berry connection whose matrix elements are $\mathcal{A}_{ab} = i \braket{u_{\v{k}a}|\partial_{\v{k}} u_{\v{k}b}}$. Note that the covariant derivative of an operator is \textit{not} the derivative of its matrix elements.

In terms of the covariant derivatives, the Hamiltonian can be written as a Taylor series in terms of the electric field as
\begin{equation}
	\widehat{H}_A = \widehat{H}_0 + \widehat{V}_E(t) = \widehat{H}_0 + \sum_{n=1}^\infty \frac{1}{n!} \left[\;\prod_{k=1}^n  \frac{e}{\hbar} A^{\alpha_k} \widehat{D}^{\alpha_k} \right] \widehat{H}_0,
	\label{eq:velocity_gauge_expansion}
\end{equation}
where $\alpha_k \in \st{x,y,z}$ is a spatial index with an implicit sum, and $\widehat{\v{D}}$ is the (Berry) covariant derivative. (Greek indices $\mu,\alpha,\beta,\dots$ will always represent spatial indices with an implicit summation henceforth.) 

Equation \eqref{eq:convariant_derivative_on_operators} can be used to write the velocity operator of the \textit{unperturbed} Hamiltonian as
\begin{equation}
	\widehat{\v{v}}  = [\widehat{\v{D}}, H_0] = -i [\widehat{\v{r}}, \widehat{H}_0].
	\label{eq:unperturbed_velocity_operator}
\end{equation}
For convenience, we define higher derivatives of the unperturbed Hamiltonian by
\begin{equation}
	\widehat{h}^{\alpha_1\dots\alpha_N} = \widehat{D}^{\alpha_1}\cdots \widehat{D}^{\alpha_N} [\widehat{H}_0].
	\label{eq:higher_derivatives_of_H}
\end{equation}

The perturbation due to the external field can then be written as
\begin{equation}
	\widehat{V}_E(t) = \frac{e}{\hbar} A^\alpha(t) \widehat{h}^\alpha + \frac{1}{2} \left( \frac{e}{\hbar} \right)^2 A^\alpha(t) A^\beta(t) \widehat{h}^{\alpha\beta}+\cdots
	\label{eq:perturbation_time_domain}
\end{equation}
Fourier transforming and using $\v{E}(\omega) = i \omega \v{A}(\omega)$, we have
\begin{equation}
	\widehat{V}_E(t) = \sum_{n=1}^\infty \frac{1}{n!} \prod_{k=1}^n \int d\omega_k e^{-i\omega_k t} \left( \frac{ie}{\hbar \omega_k} \right) E^{\alpha_k}(\omega_k) \widehat{h}^{\alpha_1\dots \alpha_n}.
	\label{eq:perturbation_frequency_domain}
\end{equation}
It is essential that---in the velocity gauge---a seemingly new perturbation appears at each order in the electric field. Physically, the $n$th term corresponds to the simultaneous interaction of $n$ photons with an electron.

The electromagnetic response of a crystal is characterized by the conductivity tensors. Incident electric fields produce a current, giving rise to a non-zero expectation of the current operator. The conductivity tensors are defined as the coefficients in an expansion of the current in powers of the external field:
\begin{align}
	\label{eq:conductivity_definition}
	&\braket{\widehat{J}^\mu}(\omega) = \int d\omega_1\; \sigma^{\mu\alpha}(\omega;\omega_1) E^\alpha(\omega_1)\\
	&+\int d\omega_1  d\omega_2\; \sigma^{\mu\alpha \beta}(\omega;\omega_1, \omega_2) E^\alpha(\omega_1) E^\alpha(\omega_2) + \cdots
\end{align}
where the first argument of the conductivity tensor $\sigma$ is the ``output'' frequency $\omega$ and the others ($\omega_1,\omega_2,\dots$) are the frequencies of the incident light.

\subsection{Feynman Rules}
The task in front of us is to compute the conductivity tensors from the Hamiltonian $\widehat{H}_0$. This is an ideal task for perturbation theory, as we start with a free fermion system and have a perturbation naturally stratified in powers of the external field. In the literature, the current operator has commonly been computed with a density matrix formalism in the single-particle picture \cite{butcher1991elements,sipe2000second}. However, we shall adopt a path-integral and Feynman diagram approach that is shorter and more physically transparent. The two approaches are, of course, equivalent.

Formally, the partition function of the perturbed system may be written as a path integral
\begin{equation}
	\begin{aligned}
		Z[\v{E}] \ &=\ \int \mathcal{D} c^\dagger \mathcal{D} c\; \exp\left( - i\int dt \, H_A(t) \right)\\
		H_A(t) &= \ibz \; c_k^\dagger(t) H_0 c_k(t) + c_k^\dagger(t) V_E(t) c_k(t).
	\end{aligned}
	\label{eq:perturbed_partition_fcn}
\end{equation}
The expectation of the current is then
\begin{align}
	\braket{\widehat{J}^\mu}(t)
	\label{eq:current_observable}
	\ &=\ \frac{1}{Z} \Tr\left[\mathcal{T} e\widehat{v}_E^\mu(t) \, e^{-i\int \widehat{H}(t') \, dt'} \right]\\
	\nonumber
	\ &=\ \frac{1}{Z} \int \mathcal{D} c^\dagger \mathcal{D} c\; \big[e v_E^\mu(t)\big] \exp\left( -i \int dt' H_A(t') \right)
\end{align}
where $\mathcal{T}$ represents time-ordering of operators. Here $\widehat{\v{v}}_E$ is the velocity operator in the \textit{perturbed} system---which itself depends on the electric field:
\begin{align}
	\label{eq:velocity_operator}
	\widehat{v}_E^\mu(t) \ &=\ \widehat{D}^\mu[\widehat{H}_0 + \widehat{V}_E(t)]\\
	\nonumber
	\ &=\ \sum_{n=0}^\infty \frac{1}{n!} \prod_{k=1}^n \int d\omega_k e^{i \omega_k t} \left( \frac{e}{\hbar \omega_k} \right) E^{\alpha_k}(\omega_k) \widehat{h}^{\mu\alpha_1\cdots \alpha_n}.
\end{align}
In terms of functional derivatives, the conductivities are then given by
\begin{align}
	\label{eq:conductivity_from_functional_deriv}
	&\sigma^{\mu\alpha_1\dots \alpha_n}(\omega; \omega_k)\\
	\nonumber
	&= \int \frac{dt}{2\pi} e^{i \omega t} \prod_{k=1}^n \int \frac{dt_k}{2\pi} e^{i \omega_k t_k} \frac{\delta}{\delta E^\alpha_k(t_k)} \braket{\widehat{J}^\mu}(t) \Big|_{\v{E}=0}.
\end{align}
As a brief technical aside, one would usually take functional derivatives in the frequency domain, but due to the explicit time-dependence in the Hamiltonian, it is neccesary to compute first in the time-domain and then Fourier transform.

Considering the form of \eqref{eq:current_observable}, we are performing a dual expansion in $\v{E}$, as both the exponent and the velocity operator depend on the electric field. As is usual in quantum field theory, the effect of the functional derivatives in \eqref{eq:conductivity_from_functional_deriv} turns out to be purely combinatorial and can be entirely captured in terms of Feynman diagrams. The only wrinkle is that, since we are computing a non-standard type of current, there is an extra vertex corresponding to the ``output'' velocity operator. 

Explicitly, the value of the $N$th non-linear conductivity can be computed by drawing all connected diagrams such that:
\begin{enumerate}
	\item There are $N+1$ external photons.
	\item All electrons are internal and compose one loop.
	\item Exactly one vertex is crossed to indicate the output current $\widehat{J}^\mu$; all other vertices are dotted.
	\item Diagrams are symmetrized over all incoming photons $(\alpha_k,\omega_k)$. The factors on the vertices in Table \ref{tab:feynman_rules} are chosen to avoid double-counting.
	\item The value of edges and vertices are given in Table \ref{tab:feynman_rules}.
\end{enumerate}
This procedure is slightly different from what is common in particle physics and thus merits some explanation. First, since $c \gg v_F$, the Fermi velocity, a negligible amount of momentum is exchanged through interactions. We thus consider only energy conservation at each vertex. Second, we assume electrons are bound to the solid, so only photons may be external. Third, since electrons must return to their equilibrium positions after a perturbation and are non-interacting, only diagrams with exactly one fermion loop are permitted. Fourth, we treat the photon as a classical background field without dynamics, whose propagator is unity. However, the electron propagator is the usual one for free fermions
\begin{equation}
	G_{\v{k}a}(\omega) = \frac{1}{\omega-\varepsilon_a(\v{k})}
	\label{eq:electron_propagator}
\end{equation}
where the $\v{k}$-index is suppressed below for notational convenience. We will see below that, in practice, photons may never cross the inside of a fermion loop.

This method is simple to apply in practice. Unlike many diagrammatic methods, this method involves no divergences beyond simple poles and does not require regularization. The computation of the first, second, and third-order responses are no more than a few lines. The only non-trivial part consists of one new contour integral at each order, which are performed for the reader's convenience in Appendix \ref{app:integrals}.

\subsection{Assumptions and Caveats}
\label{subsec:caveats}

Though the method of Feynman diagrams outlined here is convenient, it is important to recognize the assumptions that went into it and thus determine its range of validity. The use of the velocity gauge is associated with several problems: spurious divergences and inaccurate approximations. The conductivities computed in velocity gauge are apparently divergent with $\sigma^{(N)} \sim \frac{1}{\omega^N}$. These divergences are spurious, but eliminating them requires the use of sum rules. These sum rules are now understood as identities needed to convert from velocity to length gauges (see the Appendix of \cite{ventura2017gauge}). However, they are still inconvenient to apply beyond first order, so when taking the $\omega\to 0$ limit, it is best to work in the length gauge. This is carried out carefully in Section \ref{sec:semiclassics}.

The velocity gauge has often been considered badly behaved under approximations. When materials are modelled by effective Hamiltonians focusing on a few bands close to the Fermi level (such as two band models for Dirac semimetals), then effective optical responses calculated in the length gauge are generally accurate, while those in the velocity gauge can suffer from corrections the same size as the response, rendering them practically useless. It was shown in \cite{passos2017nonlinear} that this inaccuracy only arises with models where the effective Hamiltonian is not defined on the full Brillouin zone, ruining periodicity, or from the application of incorrect sum rules. In practice, this prevents some two-band models of topological materials from being studied with velocity gauge. However, if enough care is taken in defining the model, there is no reason the velocity gauge cannot be used.

One should also take care with dynamical effects. We have taken a perturbative approach to what is actually a non-equilibrium problem. The currents described here are only the initial current created after an incident pulse of light. In practice, other dynamical effects may come into play before those currents can be observed, corrupting or distorting the current. For instance, a strong laser field could create a population of excitons whose recombination interferes with the motion of electrons. This type of issue makes it difficult to observe phenomena which manifest as electrical currents rather than optical responses, such as the shift current. We should note, however, that the perturbation theory with equilibrium Green's functions accurately describe the nonlinear conductivities, since they are obtained as finite order perturbation in the external electric field $\v{E}(\omega)$ with respect to the equilibrium state. Namely, one could say that our diagrammatic approach generalizes the Kubo formula for linear response. Normally the Kubo formula relates the linear conductivity to the current-current correlation function $\braket{\v{J}\v{J}}$. Nonlinear conductivity generalizes this to the setting where the input and output current are different operators, so we are effectively computing correlators of the form $\braket{\v{J} \left( \v{J}\v{J} \v{J} \right)}$ (at the third order). 

A brief comment on the effect of scattering is also in order. In a real material, impurities, photons and other effects will perturb the free fermion band-structure. If these effects are sufficiently weak, as is often the case, one can simply replace the electron propagator with a dressed version
\begin{equation}
	G_a(\omega) = \frac{1}{\omega-\varepsilon_a} \to \frac{1}{\omega - \varepsilon_a + i\Sigma_a(\omega)}
\end{equation}
where $\Sigma_a$ is the self-energy of the electron, and is calculable within Fermi liquid theory. In practice it is usually unnecessary to understand the full frequency dependence of the self-energy. The phenomenological approximation $i\Sigma(\omega) = i \gamma \to i0^+$ is therefore often made. All the above calculations can be carried out with this phenomenological scattering factor included by slightly moving the poles, i.e. simply substituting $\omega_1 \to \omega_1 + i \gamma$, etc. One should note that for two-photon poles, the scattering factor contributes twice, so
\begin{equation}
	\frac{1}{\omega_1 + \omega_2 - \varepsilon} \to \frac{1}{\omega_1 + \omega_2 - \varepsilon - 2i\gamma}.
\end{equation}
It was pointed out in \cite{passos2017nonlinear} that this factor of two can actually have a significant effect on the shape of resonances, especially at low frequency, and is therefore crucial when making experimental predictions.

This procedure is essentially the same as incorporating interactions into the model. In principle, the technique developed here works in the fully interacting case, once the propagator and velocity operator are appropriately modified. However, performing this analytically requires either weak interactions (i.e. a Fermi liquid) or a quadratic Hamiltonian, such as in a mean-field approximation. The BCS model of superconductivity falls into the later category, and non-linear responses of superconductors will be the topic of future work.

Equipped with the Feynman rules, it is straightforward to compute the non-linear conductivity tensors at any order. At first-order there are two diagrams, four at second order, and eight at third order. Each corresponds to a unique physical process that contributes independently to the overall response. 

\section{First Order Conductivity}
\label{sec:first_order}

As a pedagogical demonstration of our framework, we re-derive the first-order conductivity. Using the rules, the answer is almost immediate. As an additional confirmation, however, we offer a complementary derivation starting from the definition of the conductivity. One can see this as a derivation of the Feynman rules at first order.

\subsection{Derivation from Diagrams}

There are two Feynman diagrams at first order:
\begin{align}
	\label{eq:feynman_diagrams_first_order}
	&\sigma^{\mu\alpha}(\omega;\omega_1) = \\
\nonumber
	&\begin{tikzpicture}[baseline=(a.center)]
		\node (a) {\includegraphics{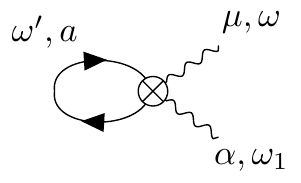}};
	\end{tikzpicture}
	+ 
	\begin{tikzpicture}[baseline=(a.center)]
		\node (a) {\includegraphics{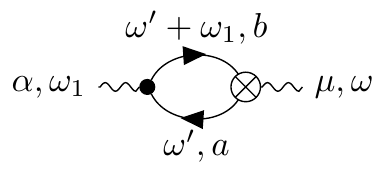}};
\end{tikzpicture}\\
\nonumber
\ &=\ 
\frac{ie^2}{\hbar \omega_1 } \sum_{a,b} \ibz \int d\omega' \; h_{ab}^\alpha G_b(\omega'+\omega_1) h_{ba}^\mu G_a(\omega')\\
\nonumber
&\hspace{1em}+ \frac{ie^2}{\hbar\omega_1}  \sum_{a} \ibz \int d\omega' \; h^{\mu\alpha}_{aa} G_a(\omega').
\end{align}
The frequency integrals are performed with standard techniques (see Appendix \ref{app:integrals}) to find
\begin{align}
	I_1 &= \int d\omega' G_a(\omega') \ =\ f_a\\
	I_2(\omega_1) &= \int d\omega' G_b(\omega'+\omega_1)  G_a(\omega') \ =\ \frac{f_{ab}}{\omega_1-{\varepsilon_{ab}}}
\end{align}
where $f_a = f(\varepsilon_{\v{k}a})$ is the Fermi factor and $f_{ab} = f_a - f_b$, $\varepsilon_{ab} = \varepsilon_{a} - \varepsilon_b$ are differences of Fermi factors and energies respectively. Therefore the conductivity is
\begin{equation}
	\sigma^{\mu\alpha}(\omega;\omega) = \frac{ie^2}{\hbar \omega}\sum_{a\neq b} \ibz f_a h^{\mu\alpha}_{aa} + \frac{h_{ab}^\alpha h_{ba}^\mu f_{ab} }{\omega-\varepsilon_{ab}}.
	\label{eq:conductivity_first_order}
\end{equation}
(The sum over band indices is only performed over the indices appearing in each term; the first term is summed over $a$ while the second is summed over both $a$ and $b$. This notational abbreviation is used from now on.)

To connect this to familiar results, we convert to the length gauge and consider the $\omega \to 0$ limit, expressing all matrix elements in terms of the velocity matrix elements, $v^\mu_{ab} = h^{\mu}_{ab}$. Using the identity
\begin{equation}
	h^{\mu\alpha}_{ab} = [D^\alpha,v^\mu]_{ab} = \partial^\alpha v^\mu_{ab} - i [A^\alpha, v^\mu]_{ab}
	\label{eq:h_2_to_velocity_matrix_elements}
\end{equation}
and the fact $v_{ab}^\mu = -i \varepsilon_{ba} A_{ab}^\mu$, the conductivity becomes
	\begin{align}
	\label{eq:conductivity_drude_form}
	&\sigma^{\mu\alpha}(\omega;\omega) =\\
	\nonumber
	&\frac{ie^2}{\hbar \omega} \sum_{a, b} \ibz f_a \partial^\alpha v_{aa}^\mu + f_{ab} v^\alpha_{ab} v^\mu_{ba} \left( \frac{1}{\varepsilon_{ba}} - \frac{1}{\omega-\varepsilon_{ab}} \right).
\end{align}
Combining the term in parentheses into a single fraction eliminates the spurious divergence:
\begin{equation}
	\sigma^{\mu\alpha}(\omega,\omega) = \frac{i e^2}{\hbar} \sum_{a, b} \frac{f_a \partial^\alpha v^\mu_{aa}}{\omega} +\frac{f_{ab} v^\alpha_{ab} v^\mu_{ba}}{(\omega-\varepsilon_{ab})\varepsilon_{ba}}
	\label{eq:conductivity_no_spurious_divergence}
\end{equation}
This is the standard result in the length gauge \cite{sipe2000second}.

In the $\omega \to 0$ limit, the second term becomes $f_{ab} v^{\alpha}_{ab} v^{\mu}_{ba}/(\varepsilon_{ba}^2) + O(\omega^2)= f_a\mathcal{F}^{\mu\alpha}_{aa} + O(\omega^2)$, the Berry curvature. We then have
\begin{align}
	\label{eq:conductivity_drude_form_SC_form}
	&\lim_{\omega \to 0} \sigma^{\mu\alpha}(\omega;\omega) =\\
	\nonumber
	&\frac{ie^2}{\hbar} \sum_{a} \ibz \frac{-\partial^\alpha f_a v_{aa}^\mu}{\omega -i\gamma} + f_{a} \mathcal{F}^{\mu\alpha}_{aa}	
\end{align}
The first term corresponds to the Drude weight, and the second term is responsible for the Hall conductivity. This formula matches what is derived from semiclassics in a Boltzmann equation approach, which is examined in Section \ref{sec:semiclassics}.

\subsection{Derivation of the Diagrams}

We now give an alternative derivation of Eq. \eqref{eq:feynman_diagrams_first_order} from the definition of the current operator. This is essentially equivalent to a derivation of the Feynman rules and may be skipped by a reader already convinced of their validity. We start from the time-domain conductivity
\begin{align}
	\sigma^{\mu\alpha}(t;t_1)
	\ &=\  \frac{\delta}{\delta E^\alpha(t_1)} \braket{\widehat{J}^\mu}(t) \Big|_{\v{E}=0}.
	\label{eq:conductivity_first_order_time_domain}
\end{align}
We must therefore evaluate the expectation value of
\begin{equation}
		\frac{\delta \widehat{v}_E^\mu(t)}{\delta E^\alpha(t_1)}
		- \widehat{v}_E^\mu(t) 
		\frac{\delta}{\delta E^\alpha(t_1)}  \int  dt' \;\widehat{H}(t') 
		\label{eq:two_terms_for_sigma_1}
\end{equation}
at $\v{E} = 0$. Writing $\widehat{H}(t') = \widehat{H}_0 + \widehat{V}_E(t')$, and using \eqref{eq:perturbation_frequency_domain}, the second term requires the derivative 
\begin{align}
	&\frac{\delta}{\delta E^\alpha(t_1)} \widehat{V}_E(t)\\
	\nonumber
	&= \frac{ie}{\hbar} \int d\omega_1 \frac{e^{-i \omega_1(t-t_1)}}{\omega_1}\\
	\nonumber
	\ &\qquad\times 
	\sum_{n=0}^\infty \frac{1}{n!} \prod_{k=1}^n \int d\omega_k e^{-i \omega_k t} E^{\alpha_k} \widehat{h}^{\alpha \alpha_1 \dots \alpha_n}\\
\nonumber
&=  \frac{ie}{\hbar} \int d\omega_1 \frac{e^{-i \omega_1(t-t_1)}}{\omega_1} \widehat{h}^\alpha + O(\v{E}). 
	\label{eq:functional_derivative_of_potential}
\end{align}
Starting from \eqref{eq:velocity_operator} for the first term of \eqref{eq:two_terms_for_sigma_1}, one computes
\begin{align}
	&\frac{\delta}{\delta E^\alpha(t_1)} \widehat{v}^{\mu}_E(t)\\ 
	\nonumber
	\ &=\ \frac{ie}{\hbar} \int d\omega_1 \frac{e^{-i \omega_1(t-t_1)}}{\omega_1}\\
	\nonumber
	\ &\qquad\times 
	\sum_{n=0}^\infty \frac{1}{n!} \prod_{k=1}^n \int d\omega_k e^{-i \omega_k t} E^{\alpha_k} \widehat{h}^{\mu \alpha \alpha_1 \dots \alpha_n}(t)\\
\nonumber
&=  \frac{ie}{\hbar} \int d\omega_1 \frac{e^{-i \omega_1(t-t_1)}}{\omega_1} \widehat{h}^{\mu\alpha}(t) + O(\v{E}). 
	\label{eq:functional_derivative_of_velocity}
\end{align}
Hence the conductivity is
\begin{align}
	&\sigma^{\mu\alpha}(t;t_1)\\
	\nonumber
	\ &=\ - e \int dt' \frac{ie}{\hbar} \int d\omega_1 \frac{e^{-i\omega_1(t'-t_1)}}{\omega_1} \braket{\widehat{h}^\mu(t) \widehat{h}^\alpha(t')}\\
	\nonumber
	\ &\hspace{2em}+\ e \frac{ie}{\hbar} \int d\omega_1 \frac{e^{-i \omega_1(t-t_1)}}{\omega_1} \braket{\widehat{h}^{\mu\alpha}(t)},
\end{align}
where brackets denote expectations with respect to the \textit{unperturbed} Hamiltonian.

To proceed, we must evaluate the expectation values in terms of the electron propagator
\begin{equation}
	\braket{c_{\v{k}a}^\dagger(t) c_{\v{k}b}(t')} = \delta_{ab} \int d\omega\; e^{i \omega(t-t')} G_{\v{k}a}(\omega).
\end{equation}
Hence
\begin{align}
	\braket{\widehat{h}^{\mu\alpha}(t)} 
	\ &=\ \sum_{a,b} \ibz \;  \braket{c^\dagger_{ka}(t) h^{\mu\alpha}_{ab} c_{kb}(t)}\\
	\ &=\ \sum_{a} \ibz \; h^{\mu\alpha}_{aa} \int d\omega\; G_{ka}(\omega)
\end{align}
and, applying Wick's theorem,
\begin{align}
	&\braket{\widehat{h}^\mu(t) \widehat{h}^\alpha(t')}\\
	\nonumber
	\ &=\ \sum_{a,b,c,d} \ibz \braket{c_{ka}^\dagger(t) h^{\mu}_{ab} c_{kb}(t) c^\dagger_{kc}(t') h^\alpha_{cd} c_{kd}(t')}\\
	\nonumber
	\ &=\ \sum_{a,b} \ibz h^\mu_{ba} G_{ka}(t-t') h^\alpha_{ab} G_{kb}(t'-t)\\
	\nonumber
	\ &=\ -\sum_{a,b} \ibz \int d\omega'' e^{-i\omega''(t_1-t')} \int d\omega' e^{-i \omega'(t'-t_1)}\\
	\nonumber
	\ &\hspace{3em} \times  h^\mu_{ba} G_{ka}(\omega'') h^\alpha_{ab} G_{kb}(\omega').
\end{align}
In the last step we dropped terms corresponding to disconnected diagrams, which contribute zero in expectation.

We have now reduced everything to the propagators and matrix elements of derivatives of the Hamiltonian---the elements present in the Feynman rules. The last step is to Fourier transform the conductivity to frequency-space to eliminate exponential factors. Thus

\begin{align}
	&\sigma^{\mu\alpha}(\omega;\omega_1)\\
	\nonumber
	&=\int \frac{dt}{2\pi} e^{i\omega t} \int \frac{dt'}{2\pi} e^{i \omega_1 t_1} \frac{ie^2}{\hbar} \left[ 
		\int d\omega_1 \frac{e^{-i\omega_1(t-t_1)}}{\omega_1} \braket{\widehat{h}^{\mu\alpha}(t)}\right. \\
	\nonumber
		& \hspace{2em} \left.
		-\int dt' \int d\omega_1 \frac{e^{-i\omega_1 (t'-t_1)}}{\omega_1} \braket{\widehat{h}^\mu(t) \widehat{h}^\alpha(t')}\ 
\right]\\
	\nonumber
\ &=\ \frac{ie^2}{\hbar \omega} \sum_{a,b} \ibz \left[ \int d\omega' \; h^{\mu\alpha}_{aa} G_a(\omega')\right.\\
	\nonumber
	&\hspace{2em} \left.+ \int d\omega'\; h^\alpha_{ab} G_a(\omega') h^{\mu}_{ba} G_b(\omega'-\omega_1) \right] \delta(\omega-\omega_1)
\end{align}
where in the first step all the $t$ integrals have been performed to create $\delta$-functions in frequency, eliminating $t,t', \omega'$ and $\omega''$ and $\omega_1 \to \omega'$ in the second step. This precisely matches \eqref{eq:feynman_diagrams_first_order}, which was obtained immediately by Feynman diagrams. The diagrams serve to eliminate the tedious steps of collapsing Fourier transforms into $\delta$ functions, thereby greatly streamlining calculations.

\section{Second Order Response}
\label{sec:second_order}

We now turn to the second-order response and demonstrate the second-order conductivity is concisely reproduced by the diagramatic formalism. There are four diagrams that contribute:
\begin{align}
	\label{eq:second_order_response}
	&\sigma^{\mu\alpha\beta}(\omega;\omega_1,\omega_2) =\\
	\nonumber
	\ &
	\begin{tikzpicture}[baseline=(a.center)]
		\node (a) {\includegraphics{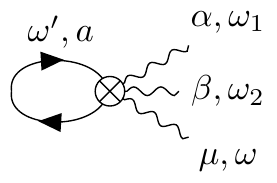}};
	\end{tikzpicture}
+
\begin{tikzpicture}[baseline=(a.center)]
\node (a) {\includegraphics{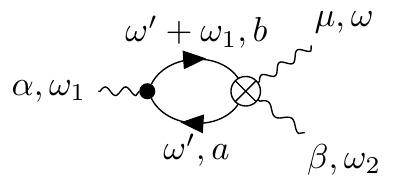}};
\end{tikzpicture}+\\
\nonumber
&
\begin{tikzpicture}[baseline=(a.center)]
		\node (a) {\includegraphics{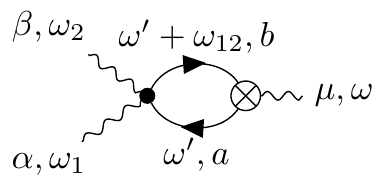}};
	\end{tikzpicture}
+
\begin{tikzpicture}[baseline=(a.center)]
		\node (a) {\includegraphics{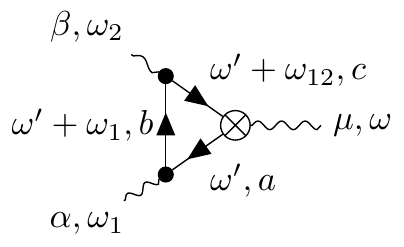}};
\end{tikzpicture}\\
\nonumber
&+ \big( (\alpha, \omega_1) \leftrightarrow (\beta, \omega_2) \big)\\[0.5em]
\nonumber
\ &=\ e\left( \frac{ie}{\hbar \omega_1} \right) \left( \frac{ie}{2\hbar \omega_2} \right) \sum_a \ibz \int d\omega' G_a(\omega') h^{\mu\alpha\beta}_{aa}\\
\nonumber
&+ \frac{-e^3}{\hbar^2 \omega_1 \omega_2} \sum_{a,b} \ibz \int d\omega' G_a(\omega') h_{ab}^\alpha G_b(\omega' + \omega_1) h^{\mu\beta}_{ba}\\
\nonumber
&+ \frac{-e^3}{2\hbar^2 \omega_1 \omega_2} \sum_{a,b} \ibz \int d\omega' G_a(\omega') h^{\alpha\beta}_{ab} G_b(\omega'+\omega_{12}) h^\mu_{ba}\\
\nonumber
&+ \frac{-e^3}{\hbar^2 \omega_1 \omega_2} \sum_{a,b,c} \ibz \int d\omega' G_a(\omega') h^\alpha_{ab} G_b(\omega'+\omega_1) h^\beta_{bc}\\
\nonumber
&\hspace{5em} \times  G_c(\omega'+\omega_{12}) h^\mu_{ca}
+ \big(( \alpha, \omega_1 ) \leftrightarrow (\beta, \omega_2) \big).
\end{align}
There is an overall constraint $\omega = \omega_{12} \equiv \omega_1 + \omega_2$.

The frequency integrals, which are called $I_1, I_2$, and $I_3$ are performed in Appendix \ref{app:integrals}. Indeed, only the triangle diagram contributes a new integral, $I_3$, computed in \eqref{eq:I_3}; the others appeared at first order. The most convenient form of $I_3$ depends on the situation. For instance, one can use partial fractions to split each term into separate resonances. However, for now we adopt a more compact representation with a triple resonance:
\begin{equation}
	\label{eq:I_3_triple_resonance}
	I_3(\omega_1,\omega_2) = 
	\frac{(\omega_2-\varepsilon_{cb}) f_{ab} + (\omega_1 - \varepsilon_{ba}) f_{cb}}{\left( \omega_1 - \varepsilon_{ba} \right)\left(  \omega_2 - \varepsilon_{cb} \right) \left( \omega - \varepsilon_{ca} \right)}
\end{equation}
We thus arrive at a formula for the second-order conductivity
\begin{align}
	\label{eq:second_order_conductivity}
	&\sigma^{\mu\alpha\beta}(\omega;\omega_1,\omega_2) =\\
	\nonumber
	&\frac{-e^3}{\hbar^2 \omega_1 \omega_2} \sum_{a,b,c} \ibz \;
	\frac{1}{2}f_a h^{\mu\alpha\beta}_{aa}\
	+ f_{ab} \frac{h^\alpha_{ab} h^{\mu\beta}_{ba}}{\omega_1 - \varepsilon_{ab}}
	+ f_{ab} \frac{\frac{1}{2}h^{\alpha\beta}_{ab} h^\mu_{ba}}{\omega - \varepsilon_{ab}}\\
	\nonumber
	&\hspace{2em}+ h^\alpha_{ab} h^\beta_{bc} h^\mu_{ca}
	\frac{(\omega_2-\varepsilon_{cb}) f_{ab} + (\omega_1 - \varepsilon_{ba}) f_{cb}}{\left( \omega_1 - \varepsilon_{ba} \right)\left(  \omega_2 - \varepsilon_{cb} \right) \left( \omega - \varepsilon_{ca} \right)}\\
	\nonumber
	&\hspace{2em}+ \big[ ( \alpha, \omega_1 ) \leftrightarrow (\beta, \omega_2) \big].
\end{align}
As above, the sum over bands $a,b,c$ should only be employed when necessary. For instance, the term $f_a h^{\mu\alpha\beta}_{aa}$ is only summed over $a$, and not $b,c$.

Let us pause for a moment to interpret the structure of this formula. Each term is a product of a matrix-element part and a resonance part from one of the integrals $I_1, I_2$ or $I_3$. This natural separation allows us to easily consider various physical limits, wherein the resonance structure simplifies but the matrix elements remain unchanged. The terms are arranged by powers of $\omega$. The first term corresponds to the derivative of the Drude weight, the ``Drude weight dipole''. The second and third terms are one- and two-photon resonances respectively, which are large when two bands are separated by energies of $\omega_1$ or $\omega_1+\omega_2$. The last term, corresponding to the triangle diagram, is more complex. We will see below that it is still the sum of one-photon and two-photon resonances.\footnote{However, this term is familiar from atomic physics: the second-order response of a molecular system has the same form as this last term, but without the Brillouin zone integral. Of course, the meaning of the matrix elements is different in that situation.} Also note that there is an overall pole $(\omega_1 \omega_2)^{-1}$. Except for the first term, the resonance factors can be used to eliminate this apparent divergence. The exception is in the $\omega \to 0$ limit, which contains a physical divergence.	Section \ref{sec:semiclassics} considers this point carefully.

To provide convenient equations for important limits, as well as to gain a better understanding of the resonance structure, we next examine two limits: second-harmonic generation and the shift current. Again, this merely amounts to taking the limit of the resonance integrals $I_2$ and $I_3$, and other limits can be carried out with comparable ease.

\subsection{Second-Harmonic Generation}

The second-harmonic response is generated by both one-photon and two-photon resonances. That is, if the incident light is at frequency $\omega$, then there will be a second-order response at both $\omega$ and $2\omega$. One-photon resonances come from the second and fourth diagrams in \eqref{eq:second_order_response}, while two-photon resonances are due to the third and fourth diagrams. The first diagram only contributes resonantly near $\omega = 0$. To capture these resonances carefully, we use an alternative form for the integral $I_3$ which makes them manifest.

Defining $\rho_1 = \omega_1/\omega, \rho_2 = \omega_2/\omega$. We may apply the partial fraction identity
\begin{align}
	&\frac{1}{(A-\omega_1)(B-\omega_{2})} =\\
	\nonumber
	&\hspace{4em} \frac{1}{(A-\rho_1 B)\left( B-\omega_{2} \right)} - \frac{\rho_1}{(A-\rho_1 B)(A-\omega_1)}
\end{align}
to write
\begin{align}
\label{eq:I_3_separate_poles}
		&I_3(\omega_1,\omega_2)\\ 
		\nonumber
		\ &=\ \frac{1}{\rho_1 \varepsilon_{cb} + \rho_2 \varepsilon_{ab}}\left[ \frac{f_{ac}}{\omega - \varepsilon_{ca}} + \frac{\rho_2 f_{bc}}{\omega_2 - \varepsilon_{ab}} + \frac{\rho_1 f_{ba}}{\omega_1 - \varepsilon_{ba}}\right].
\end{align}

Here the first-term is a resonance due to absorbing both photons simultaneously, while the latter two are resonances in $\omega_1$ or $\omega_2$ only. So far, this is general, and can be used in \eqref{eq:second_order_conductivity} in place of \eqref{eq:I_3_triple_resonance}. In the case of second harmonic generation, we take $\omega_1 = \omega_2 = \omega$, so $\rho_1 = \rho_2 = \frac{1}{2}$. After several cancellations,
\begin{align}
	&I_3(\omega,\omega)\\
	\nonumber
	&= \frac{1}{\varepsilon_{ab} + \varepsilon_{cb}}
	\left[ \frac{2f_{ac}}{2\omega - \varepsilon_{ca}} 
			+ \frac{f_{bc}}{\omega - \varepsilon_{cb}}
			+ \frac{f_{ba}}{\omega - \varepsilon_{ba}} 
\right].
\label{eq:I_3_SHG_limit}
\end{align}

Starting from the general equation \eqref{eq:second_order_conductivity}, using $I_3(\omega,\omega)$, and writing out the frequency-symmetrization $(\alpha,\omega) \leftrightarrow (\beta, \omega)$ yields
\begin{align}
	&\sigma^{\mu\alpha\beta}(2\omega;\omega,\omega)
	 =\\
	 \nonumber
	&- \frac{e^3}{2\hbar^2 \omega^2} \sum_{a,b,c} \ibz \;
	f_a h^{\mu\alpha\beta}_{aa}
	+ f_{ab} \frac{h^\alpha_{ab} h^{\mu\beta}_{ba} + h^\beta_{ab} h^{\mu\alpha}_{ba}}{\omega - \varepsilon_{ab}} \\
	\nonumber
	&\hspace{4em}+ f_{ab} \frac{h^{\alpha\beta}_{ab} h^\mu_{ba}}{2\omega - \varepsilon_{ab}}
	+ \frac{ \left(h^\alpha_{ab} h^\beta_{bc} + h^\alpha_{ab} h^\beta_{bc}\right) h^\mu_{ca}}{\varepsilon_{ab} + \varepsilon_{cb}}\\
	\nonumber
	&\hspace{4em}\times 
	\left[ \frac{2f_{ac}}{2\omega - \varepsilon_{ca}} 
			+ \frac{f_{bc}}{\omega - \varepsilon_{cb}}
			+ \frac{f_{ba}}{\omega - \varepsilon_{ba}}
	\right].
\end{align}
This result is equivalent to velocity-gauge formulas for the second-harmonic present in the literature\cite{passos2017nonlinear, Yang17}, but did not involve any sum rules.

\subsection{Shift Current}
Another interesting limit to consider is the so-called shift current, $\sigma^{\mu\alpha\beta}(0;\omega,-\omega)$. It can be thought of as the ``solar panel'' response where incident light generates a DC current, and has been of recent interest in the context of two-band systems where it has a particularly simple form \cite{Morimoto-Nagaosa16}.

As with the second harmonic, the only real task is to determine what happens to the pole structure. Starting from \eqref{eq:I_3_separate_poles}, one finds
\begin{equation}
	I_3(\omega,-\omega) = \frac{1}{\varepsilon_{ac}} \left[\frac{f_{ab}}{\left( \omega - \varepsilon_{ba} \right)}
	- \frac{f_{cb}}{\left( \omega - \varepsilon_{bc} \right)} \right].
\end{equation}
Then, symmetrizing explicitly,for the prime case of interest $\alpha = \beta$
\begin{align}
	\label{eq:shift_current}
	&\sigma^{\mu\alpha\alpha}(0;\omega,-\omega) =\\
	\nonumber
	&\hspace{2em}\frac{e^3}{\hbar^2 \omega^2}
	\sum_{a,b,c} \ibz \;
	f_a h_{aa}^{\mu\alpha\alpha} 
	+ f_{ab} \frac{h_{ab}^\alpha h^{\mu\alpha}_{ba}}{\omega - \varepsilon_{ab}}\\
	\nonumber
	&\hspace{3em}+ f_{ab} \frac{h_{ab}^\alpha h^{\mu\alpha}_{ba}}{-\omega - \varepsilon_{ab}}
	+ f_{ab} \frac{h^{\alpha\alpha}_{ab} h^\mu_{ba}}{\varepsilon_{ba}}\\
	\nonumber
	&\hspace{3em}+ \frac{h^\alpha_{ab} h^\alpha_{bc} h^\mu_{ca}}{\varepsilon_{ac}}\left[ \frac{f_{ab}}{\left( \omega - \varepsilon_{ba} \right)}
	- \frac{f_{cb}}{\left( \omega - \varepsilon_{bc} \right)} \right]\\
	\nonumber
	&\hspace{3em}+ \frac{h^\alpha_{ab} h^\alpha_{bc} h^\mu_{ca}}{\varepsilon_{ac}}\left[ \frac{f_{ab}}{\left( -\omega - \varepsilon_{ba} \right)}
	- \frac{f_{cb}}{\left( -\omega - \varepsilon_{bc} \right)} \right].
\end{align}

This result agrees with known expressions for the shift current found in the literature. One can easily check this reduces to the correct two-band limit that has been studied in previous work~\cite{Morimoto-Nagaosa16}. It is worth contrasting this result to the alternative (but equivalent) length-gauge results in, e.g. Ref.~\cite{sipe2000second}. The results there involve a maximum of two bands in each term, whereas here there are three band terms. Converting between the two gauges requires the use of sum rules, which exchange some intraband matrix elements with interband ones, and visa versa. Specifically, one can convert to the shift current formula in \cite{sipe2000second} by focusing on the interband resonance from the band $a$ to $b$, where we collect terms involving $\tfrac{f_{ab}}{\left( \omega - \varepsilon_{ba} \right)}$ after switching indices $a \leftrightarrow c$ in the term $\tfrac{f_{cb}}{\left( \omega - \varepsilon_{bc} \right)}$ and use the second-order sum rule in Eq. (13) of \cite{Cook17}. Moreover, since the conductivity does not depend on the choice of gauge, one may conclude that saying a particular term involves a certain number of bands is gauge-dependent information and therefore not necessarily physical. This demonstates Eq. \eqref{eq:shift_current} is equivalent to previously known expressions for the shift current in the literature.

The injection current is a second-order process that describes a current whose magnitude grows linearly in time as the sample is illuminated. This process is manifest in in Ref.~\cite{sipe2000second} as a term with an overall $1/\delta\omega$ divergence in the nonlinear conductivity $\sigma(\delta\omega;\omega+\delta\omega,-\omega)$. In our framework, the naive limit of Equation \eqref{eq:shift_current} does not contain this divergence, but it can be recovered by considering the limit $\delta \omega \to 0$ of  $\sigma(\delta\omega;\omega+\delta\omega,-\omega)$ in Eq.~\eqref{eq:second_order_conductivity}.
Specifically, the last two terms in Eq.~\eqref{eq:shift_current} are related to the injection current since the factor $1/\epsilon_{ac}$ is divergent for $a=c$ which is cutoff by introducing a small $\delta \omega$. The subtle limit $\delta \omega \to 0$, particularly in the case $\alpha \neq \beta$, will be the subject of future work.

\bigskip

\section{Third Order Response}
\label{sec:third_order}

It is at third order that the diagrammatic method espoused here becomes the most useful.  Unlike at second-order, the third order response is generically allowed by symmetry and expected to be present to some degree in all materials. Third order optical responses are relatively unstudied, especially in the case of non-zero Berry connection. Understanding this area is our main focus in this paper. Using our diagrammatic formalism, we derive expressions for the third-order response where each term is associated with an individual process. This allows them to be split into manifestly one-photon, two-photon, and three-photons parts, so that the origin of each resonance is clear. We then examine the limits of third harmonic generation and self-focusing of light. In subsequent sections we begin to interpret these formulas in the one band (semiclassical) and two-band (Weyl semimetal) limits.

There are eight diagrams that contribute at third order.
\begin{widetext}
\begin{align}
&	\sigma^{\mu\alpha\beta\gamma}(\omega;\omega_1,\omega_2,\omega_3) =\\ 
&\begin{tikzpicture}[baseline=(current bounding box.center)]
	\node {\includegraphics{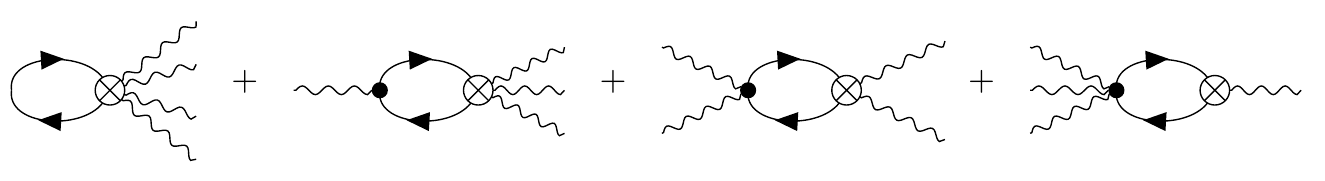}};
\end{tikzpicture}\\
&\begin{tikzpicture}[baseline=(current bounding box.center)]
	\node {\includegraphics{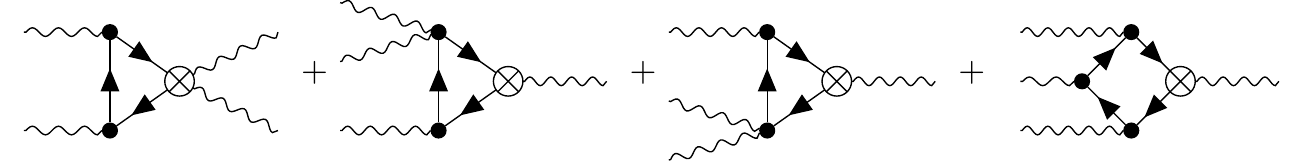}};
\end{tikzpicture}\\
&=
\frac{e}{3!\omega_1 \omega_2 \omega_3} \left( \frac{ie}{\hbar} \right)^3
\sum_a \ibz \int d\omega' \; G_a(\omega') h^{\mu\alpha\beta\gamma}_{aa}\\
&\hspace{2em}+\frac{e}{2!\omega_1 \omega_2 \omega_3} \left( \frac{ie}{\hbar} \right)^3 \sum_{a,b} \ibz \int d\omega'
\; G_a(\omega') h^\alpha_{ab} G_b(\omega' + \omega_1) h^{\mu\beta\gamma}_{ba}\\
&\hspace{2em}+\frac{e}{2!\omega_1 \omega_2 \omega_3} \left( \frac{ie}{\hbar} \right)^3 \sum_{a,b} \ibz \int d\omega'\;
G_{a}(\omega') h^{\alpha\beta}_{ab} G_b(\omega'+\omega_{12}) h^{\mu\gamma}_{ba}\\
&\hspace{2em}+\frac{e}{3!\omega_1 \omega_2 \omega_3} \left( \frac{ie}{\hbar} \right)^3 \sum_{a,b} \ibz \int d\omega'\;
G_a(\omega') h^{\alpha\beta\gamma}_{ab} G_a(\omega'+\omega_{123}) h^\mu_{ba}\\
&\hspace{2em}+\frac{e}{\omega_1 \omega_2 \omega_3} \left( \frac{ie}{\hbar} \right)^3 \sum_{a,b,c} \ibz \int d\omega'\;
G_{a}(\omega') h^{\alpha}_{ab} G_b(\omega'+\omega_1) h^\beta_{bc} G_c(\omega'+\omega_{12}) h^{\mu\gamma}_{bc}\\
&\hspace{2em}+\frac{e}{2!\omega_1 \omega_2 \omega_3} \left( \frac{ie}{\hbar} \right)^3 \sum_{a,b,c} \ibz \int d\omega'\;
G_a(\omega') h^\alpha_{ab} G_a(\omega'+\omega_1) h^{\beta\gamma}_{bc} G_{c}(\omega'+\omega_{123}) h^\mu_{ca}\\
&\hspace{2em}+\frac{e}{2!\omega_1 \omega_2 \omega_3} \left( \frac{ie}{\hbar} \right)^3 \sum_{a,b,c} \ibz \int d\omega'\;
G_a(\omega') h^{\alpha\beta}_{ab} G_a(\omega'+\omega_{12}) h^{\gamma}_{bc} G_{c}(\omega'+\omega_{123}) h^\mu_{ca}\\
&\hspace{2em}+\frac{e}{\omega_1 \omega_2 \omega_3} \left( \frac{ie}{\hbar} \right)^3 \sum_{a,b,c,d} \ibz \int d\omega'\;
G_a(\omega') h^\alpha_{ab} G_b(\omega'+\omega_1) h^\beta_{bc} G_c(\omega'+\omega_{12}) h^\gamma_{cd} G_d(\omega'+\omega_{123}) h^\mu_{da}
\end{align}
The $\omega'$ integrals are evaluated in Appendix \ref{app:integrals}. For concision, however, we shall leave the expression in terms of $I_3$ and $I_4$. We must also symmetrize under all possible combinations of incoming photons, which amounts to the six permutations of $(\alpha,\omega_1)$, $(\beta,\omega_2)$, and $(\gamma,\omega_3)$. We will denote this permutation symmetry by $\frac{1}{3!} S_3$. The full third-order non-linear response is thus
\begin{equation}
	\begin{aligned}
		&\sigma^{\mu\alpha\beta\gamma}(\omega;\omega_1,\omega_2,\omega_3) =\\
		&\hspace{2em}\frac{1}{3!} S_3 \frac{-i e^4}{\hbar^3 \omega_1 \omega_2 \omega_3} \sum_{a,b,c,d} \ibz\;
		\frac{1}{6}f_a h^{\mu\alpha\beta\gamma}_{aa}
		+ \frac{\frac{1}{2}f_{ab} h^\alpha_{ab} h^{\mu\beta\gamma}_{ba}}{\omega_1 - \varepsilon_{ab}}
		+ \frac{\frac{1}{2}f_{ab} h^{\alpha\beta}_{ab} h^{\mu\gamma}_{ba}}{\omega_{12} - \varepsilon_{ab}}
		+  \frac{\frac{1}{6}f_{ab} h^{\alpha\beta\gamma}_{ab} h^\mu_{ba}}{\omega-\varepsilon_{ab}}\\
		&\hspace{2em}
		+h^{\alpha}_{ab} h^\beta_{bc} h^{\mu\gamma}_{bc} I_3(\omega_1,\omega_2)
		+\frac{1}{2}h^{\alpha\beta}_{ab} h^{\gamma}_{bc} h^{\mu}_{bc} I_3(\omega_{12},\omega_{3})
		+\frac{1}{2}h^{\alpha}_{ab} h^{\beta\gamma}_{bc} h^{\mu}_{bc} I_3(\omega_1,\omega_{23})
		+h^{\alpha}_{ab} h^\beta_{bc} h^\gamma_{cd} h^\mu_{da} I_4(\omega_1,\omega_2,\omega_3).
	\end{aligned}
\end{equation}

\subsection{Third-Harmonic Generation}

One physical limit of interest is third-harmonic generation, when there is a single incoming frequency. There are many simplifications in this case, giving rise to a relatively simple expression. In particular, the integral for the box diagram, $I_4$, can be separated into one-, two- and three-photon resonances as
\begin{equation}
	\begin{aligned}
	I_4(\omega,\omega,\omega) &= 
	\frac{f_{ab}}{(\omega-\varepsilon_{ba})(\varepsilon_{ab} + \varepsilon_{cb})(2\varepsilon_{ab} + \varepsilon_{db})}
	+ \frac{f_{bc}}{(\omega-\varepsilon_{cb})(\varepsilon_{ab} + \varepsilon_{cb})(\varepsilon_{bc} + \varepsilon_{dc})}
	+ \frac{f_{dc}}{(\omega - \varepsilon_{dc})(\varepsilon_{cb} + \varepsilon_{cd} )(2\varepsilon_{dc} + \varepsilon_{ac})}\\
	&+ \frac{4f_{db}}{(2\omega -\varepsilon_{db})(2\varepsilon_{ab} + \varepsilon_{db})(\varepsilon_{bc} + \varepsilon_{dc})}
	+ \frac{4f_{ca}}{(2\omega - \varepsilon_{ca})(\varepsilon_{cb} + \varepsilon_{ab})(2\varepsilon_{dc} + \varepsilon_{ac})}
	+ \frac{9 f_{da}}{(3 \omega - \varepsilon_{da})(2\varepsilon_{ab} + \varepsilon_{db})(2\varepsilon_{dc} + \varepsilon_{ac})}.
\end{aligned}
	\label{eq:I_4_THG}
\end{equation}
We can similarly decompose the integrals for the triangle diagrams. The case $I_3(\omega,\omega)$ is given in Equation \eqref{eq:I_3_SHG_limit}. Similarly,
\begin{align}
	I_3(\omega,2\omega)
	\ &=\ \frac{1}{2\varepsilon_{ab} + \varepsilon_{cb}}
	\left[ \frac{3f_{ac}}{3 \omega - \varepsilon_{ca}} 
		+ \frac{2f_{cb}}{2\omega - \varepsilon_{cb}}
		+ \frac{f_{ba}}{\omega - \varepsilon_{ba}}
	\right]\\
	I_3(2\omega,\omega)
	\ &=\
	\frac{1}{2\varepsilon_{cb} + \varepsilon_{ab}}
	\left[ 
		\frac{3 f_{ac}}{3\omega - \varepsilon_{ca}}
		+ \frac{2f_{ba}}{2\omega - \varepsilon_{ba}}
		+ \frac{f_{cb}}{\omega-\varepsilon_{cb}}
	\right].
	\label{eq:I_3_THG_limits}
\end{align}

Combining these and applying the permutation symmetry yields
\begin{align}
	\label{eq:third_harmonic_generation}
	&\sigma^{\mu\alpha\beta\gamma}(3\omega;\omega,\omega,\omega) = \\
	&\frac{-ie^4}{\hbar^3 \omega^3} \sum_{a,b,c,d} \ibz\;
	f_a h^{\mu\alpha\beta\gamma}_{aa}
	+ f_{ab} \left[\frac{
			h^\alpha_{ab} h^{\mu\beta\gamma}_{ba}
			+h^\beta_{ab} h^{\mu\gamma\alpha}_{ba}
			+h^\gamma_{ab} h^{\mu\alpha\beta}_{ba}
}{\omega-\varepsilon_{ab}}
\right]
+f_{ab} \left[ \frac{
	h^{\alpha\beta}_{ab} h^{\mu\gamma}_{ba}
	+h^{\beta\gamma}_{ab} h^{\mu\beta}_{ba}
	+h^{\gamma\alpha}_{ab} h^{\mu\alpha}_{ba}
}{2\omega - \varepsilon_{ab}} \right]\\
& \hspace{2em}+  \left[
\left( h^\alpha_{ab} h^\beta_{bc} + h^\beta_{ab} h^\alpha_{bc} \right) h^{\mu\gamma}_{ca} +
\left( h^\beta_{ab} h^\gamma_{bc} + h^\gamma_{ab} h^\beta_{bc} \right) h^{\mu\beta}_{ca} +
\left( h^\gamma_{ab} h^\alpha_{bc} + h^\alpha_{ab} h^\gamma_{bc} \right) h^{\mu\alpha}_{ca} 
\right]
I_3(\omega,\omega)\\
& \hspace{2em}+ \left[ 
	h^{\alpha}_{ab} h^{\beta\gamma}_{bc} h^\mu_{ca}
	+h^{\beta}_{ab} h^{\gamma\alpha}_{bc} h^\mu_{ca}
	+h^{\gamma}_{ab} h^{\alpha\beta}_{bc} h^\mu_{ca}
\right] \Big( I_3(\omega,2\omega) + I_3(-\omega,-2\omega) \Big)\\
& \hspace{2em}+
\left[
	h^{\alpha}_{ab} h^{\beta}_{bc} h^\gamma_{cd}
+	h^{\alpha}_{ab} h^{\gamma}_{bc} h^\beta_{cd}
+	h^{\beta}_{ab} h^{\gamma}_{bc} h^\alpha_{cd}
+	h^{\beta}_{ab} h^{\alpha}_{bc} h^\gamma_{cd}
+	h^{\gamma}_{ab} h^{\alpha}_{bc} h^\beta_{cd}
+	h^{\gamma}_{ab} h^{\beta}_{bc} h^\alpha_{cd}
\right] h^\mu_{da} I_4(\omega,\omega,\omega).
\end{align}

\subsection{Self-Focusing}

Another common third-order response is the self-focusing of light, which is the modification to the linear conductivitiy due to nonlinear effects. For instance, the process wherein an excited electron absorbs photons of energy $\omega$ and then $-\omega$,
\begin{equation}
\begin{tikzpicture}[baseline=(current bounding box.center)]
	\node (a) {\includegraphics{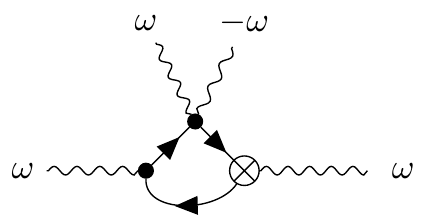}};
\end{tikzpicture}
\label{eq:self_focusing_diagram}
\end{equation}
can masquarade as the diagram for first order conductivity from Equation \eqref{eq:feynman_diagrams_first_order}. To describe this effect, one can define the effective conductivity, via $\braket{J^\mu}(\omega) = \sigma^{\mu\alpha}_{\text{eff}}(\omega) E^\alpha(\omega)$ where
\begin{equation}
	\sigma^{\mu\alpha}_\text{eff}(\omega) = \sigma^{\mu\alpha}(\omega) + \sigma^{\mu\alpha\beta\gamma}(\omega;\omega,-\omega,\omega) E^{\beta}(-\omega) E^\gamma(\omega) + O(E^4).
	\label{eq:effective_conductivity}
\end{equation}
The third-order correction term, $\sigma^{\mu\alpha\beta\gamma}(\omega;\omega,-\omega,\omega)$, is also called the self-focusing effect. 

In the self-focusing limit, the conductivity is a sum of resonances at $0\omega,1\omega$ and $2\omega$, corresponding to the sums and differences of the incident frequencies. Unfortunately, the minus sign from the $-\omega$ photons lifts the permutation symmetry between the various incident photons, creating a more complex resonance structure than in the third-harmonic case. It is convenient to express the conductivity in terms of the following expressions:
\begin{align}
	I_3(\omega,-\omega) = \frac{1}{\varepsilon_{ac}}\left[ \frac{f_{ab}}{\omega-\varepsilon_{ba}} + \frac{f_{bc}}{\omega-\varepsilon_{bc}} \right],\quad
	I_3(0,\omega) = \frac{1}{\varepsilon_{ab}}\left[ \frac{f_{ac}}{\omega-\varepsilon_{ca}} + \frac{f_{cb}}{\omega-\varepsilon_{cb}} \right],\quad
	I_3(\omega,0) = \frac{1}{\varepsilon_{bc}}\left[ \frac{f_{ab}}{\omega-\varepsilon_{ba}} + \frac{f_{ca}}{\omega-\varepsilon_{ca}} \right],
\end{align}
and $I_3(\omega,\omega)$, which is given by \eqref{eq:I_3_SHG_limit}. For the box diagram, one must also consider
\begin{align}
	&I_4(-\omega,\omega,\omega)\\
	\nonumber
	\ &=\ 
	\frac{1}{\varepsilon_{ac} (\varepsilon_{ab}+\varepsilon_{ad})}
	\left[ 
		\frac{f_{ba}}{\omega-\varepsilon_{ab}}
		+ \frac{f_{ad}}{\omega-\varepsilon_{da}}
	\right]
	+ \frac{1}{\varepsilon_{ac} (\varepsilon_{cb}+ \varepsilon_{cd})}
	\left[ 
		\frac{f_{cb}}{\omega-\varepsilon_{dc}}
		+ \frac{f_{cd}}{\omega-\varepsilon_{dc}}	
	\right]
	+ \frac{4f_{db}}{\left( 2\omega-\varepsilon_{db} \right) (\varepsilon_{bc} + \varepsilon_{dc}) (\varepsilon_{ab} + \varepsilon_{ad})},\\
	&I_4(\omega,-\omega,\omega)
	\ =\
	\frac{1}{\varepsilon_{ac} \varepsilon_{bd}}\left[ \frac{f_{ab}}{\omega-\varepsilon_{ba}} + \frac{f_{bc}}{\omega-\varepsilon_{bc}} + \frac{f_{da}}{\omega-\varepsilon_{da}} + \frac{f_{cd}}{\omega-\varepsilon_{dc}} \right]\\
	&I_4(\omega,\omega,-\omega)\\
	\nonumber
	\ &=\  \frac{1}{\varepsilon_{db} (\varepsilon_{ad} + \varepsilon_{ac})} \left[ \frac{f_{cd}}{\omega-\varepsilon_{cd}} + \frac{f_{ad}}{\omega-\varepsilon_{da}} \right]
	+
	\frac{1}{\varepsilon_{bd} \left( \varepsilon_{ab}+\varepsilon_{ac} \right)}\left[ 
		\frac{f_{ba}}{\omega-\varepsilon_{ba}}
		+ \frac{f_{cb}}{\omega-\varepsilon_{cb}}	
	\right]
	+ \frac{4 f_{ac}}{\left( 2\omega - \varepsilon_{ca} \right)\left( \varepsilon_{ab} + \varepsilon_{cb} \right)\left( \varepsilon_{ad} + \varepsilon_{cd} \right)}.
\end{align}
Then, applying the symmetrization over $(\alpha,\omega), (\beta,-\omega), (\gamma,\omega)$, the self-focusing is
\begin{align}
	&\sigma^{\mu\alpha\beta\gamma}(\omega;\omega,-\omega,\omega)=\\
\nonumber &\frac{ie^4}{3!\hbar^3 \omega^3} \sum_{a,b,c,d} \ibz f_a h_{aa}^{\mu\alpha\beta\gamma}
+ f_{ab}\left[\frac{ h^\alpha_{ab} h^{\mu\beta\gamma}_{ba}}{\omega-\varepsilon_{ab}}
	+ \frac{ h^\beta_{ab} h^{\mu\gamma\alpha}_{ba}}{(-\omega) + \varepsilon_{ab}}
	+ \frac{ h^{\gamma}_{ab} h^{\mu\alpha\beta}_{ba}}{\omega -\varepsilon_{ab}}\right]\\
	&\nonumber +f_{ab}\left[ 
	\frac{h^{\alpha\beta}_{ab} h^{\mu\gamma}}{0-\varepsilon_{ab}}
	+ \frac{h^{\beta\gamma}_{ab} h^{\mu\alpha}_{ba}}{0-\varepsilon_{ab}}
	+ \frac{h^{\gamma\alpha}_{ab} h^{\mu\beta}_{ba}}{2\omega-\varepsilon_{ab}}
\right]
+ f_{ab} \frac{h^{\alpha\beta\gamma}_{ab} h^\mu_{ba}}{\omega-\varepsilon_{ab}}\\
\nonumber
&+\left[
\left(	h^\alpha_{ab} h^{\beta}_{bc} h^{\mu\gamma}_{ca} + h^{\gamma}_{ab} h^\beta_{bc} h^{\mu\alpha}_{ca} \right) I_3(\omega,-\omega)
+\left(	h^\beta_{ab} h^{\alpha}_{bc} h^{\mu\gamma}_{ca} + h^{\beta}_{ab} h^{\gamma}_{bc} h^{\mu\alpha}_{ca} \right) I_3(-\omega,\omega)
+\left(	h^{\gamma}_{ab} h^{\alpha}_{bc} h^{\mu\beta}_{ca} + h^{\alpha}_{ab} h^{\gamma}_{bc} h^{\mu\beta}_{ca} \right) I_3(\omega,\omega)
\right]\\
\nonumber
&+\left[ 
\left(	h^{\alpha\beta}_{ab} h^\gamma_{bc} h^\mu_{ca} + h^{\beta\gamma}_{ab} h^\alpha_{bc} h^\mu_{ca} \right) I_3(0,\omega)
	+ h^{\gamma\alpha}_{ab} h^\beta_{bc} h^\mu_{ca} I(2\omega,-\omega)
\right]\\
\nonumber
&+\left[
	\left( h^{\alpha}_{ab} h^{\beta\gamma}_{bc} h^\mu_{ca} + h^{\gamma}_{ab} h^{\alpha\beta}_{bc} h^\mu_{ca} \right) I(\omega,0) + h^\beta_{ab} h^{\gamma\alpha}_{bc} h^\mu_{ca} I(-\omega,2\omega)
\right]\\
\nonumber
&+\left[ 
	\left(h^{\alpha}_{ab} h^{\beta}_{bc} h^{\gamma}_{cd} + h^{\gamma}_{ab} h^{\beta}_{bc} h^\alpha_{cd}\right) h^\mu_{da} I_4(\omega,-\omega,\omega)
+	
	\left(h^{\beta}_{ab} h^{\gamma}_{bc} h^{\alpha}_{cd} + h^{\beta}_{ab} h^{\alpha}_{bc} h^{\gamma}_{cd}\right) h^\mu_{da} I_4(-\omega,\omega,\omega)\right.\\
	\nonumber
	&\hspace{2em}+	
	\left.
	\left(h^{\gamma}_{ab} h^{\alpha}_{bc} h^{\beta}_{cd} + h^{\alpha}_{ab} h^{\gamma}_{bc} h^{\beta}_{cd}\right) h^\mu_{da} I_4(-\omega,\omega,\omega)
\right]
\end{align}
Note that there is an exact permutation symmetry $\alpha \leftrightarrow \gamma$ since the second and forth frequencies are both $\omega$.
\end{widetext}

\section{Semiclassical Limit}
\label{sec:semiclassics}

This section carefully examines the semiclassical limit of non-linear optical responses. This crucial physical limit, where on focusing on independent bands in the limit $\omega \to 0$, has been the subject of much recent work, as described in the introduction. The goal of this section is to carefully take this limit. Per the discussion in Section \ref{subsec:caveats}, this is most easily carried out in the length gauge. The alternative is to start from the velocity gauge and apply many sum rules. However, the source of these sum rules is expanding the change of gauge which converts from velocity to length gauge\cite{ventura2017gauge}, so it clear that the length gauge is the natural physical setting for this limit. We will start with a purely semiclassical derivation, then show that this matches the results from the length gauge, and lastly comment on the topological properties of the third-order semiclassical conductivity.

\subsection{Semiclassical Derivation}

We work with a single band and ignore interband contributions. Recall that the equations for semi-classical electron dynamics in an electric field (but no magnetic field) are given by
\begin{align*}
	\hbar \frac{d}{dt} \v{r} \ &=\ \v{\nabla}_{\v{k}} \varepsilon_{\v{k}} + e\v{E} \times \v{\Omega}(\v{k}),\\
	\hbar\frac{d}{dt}\v{k} \ &=\ -e\v{E}
\end{align*}
where $\v{E} = \v{E}(t)$ is the applied electric field and $\v{\Omega}$ is the standard vector representation for the Berry curvature in three dimensions. In the notation of this paper, for a single band $a$, $\Omega^\alpha_a = \varepsilon^{\alpha\beta \gamma} \mathcal{F}^{\beta\gamma}_{aa}$ where $\varepsilon$ is the Levi-Civita symbol, so $\left( \v{E} \times \v{\Omega} \right)^\mu = \mathcal{F}^{\mu \alpha} E_\alpha$. 

We take a Boltzmann equation approach, writing the charge and current density as, respectively,
\begin{equation}
	\rho(t) = -i \ibz f(t), \text{ and } \v{J}(t) = -e \ibz \frac{d\v{r}}{dt} f(t)
\end{equation}
where $f = f(t,\v{k})$ is the distribution function of electrons, and is taken to be Fermi-Dirac distribution $f_\text{FD}$ in equilibrium. The time-evolution of $f$ is given by the Boltzmann Equation
\begin{equation}
	\frac{d\v{k}}{dt} \cdot \v{\nabla}_{\v{k}} f + \partial_t f = \frac{f_\text{FD} - f}{\tau}
	\label{eq:Boltzmann_equation}
\end{equation}
for some relaxation time $\tau$. 

We take a monochromatic perturbation $\v{E}(t) = E^\alpha \v{e}_\alpha e^{i\omega t}$. Expanding $f(t) = \sum_{K \in \Z} f^{(K)} e^{-i\omega K t}$ and equating terms of the same order in \eqref{eq:Boltzmann_equation}, we have
\begin{equation}
	-e \v{E}\cdot \v{\nabla}_{\v{k}} f^{(K)} + (-i K \omega) f^{(K+1)} = -\frac{1}{\tau} f^{(K+1)}.
\end{equation}
With the initial condition $f^{(0)} = f_\text{FD}$, this gives an order-by-order solution as
\begin{equation}
	f^{(K+1)} = \frac{-i}{K\omega +i\gamma} e \v{E} \cdot \v{\nabla} f^{(K)}
	\label{eq:Boltzmann_solution}
\end{equation}
where $\gamma = 1/\tau$ is a dissipation rate. The first-order current is thus
\begin{equation}
	\v{J}^\mu(\omega) = -e \ibz \; v^\mu f^{(1)} + F^{\mu\alpha} E_\alpha f^{(0)}
\end{equation}
so, integrating by parts, the linear conductivity is
\begin{equation}
	\sigma^{\mu \alpha}(\omega;\omega) = \frac{e^2}{\hbar} \ibz \; f_\text{FD} \left( -i\frac{\partial^\alpha v^\mu}{\omega + i \gamma} - F^{\mu\alpha} \right)	
	\label{eq:semi_classical_linear_conductivity}
\end{equation}
One can check this exactly reproduces the fully quantum equation for $\sigma^{\mu\alpha}$, Equation \eqref{eq:conductivity_drude_form_SC_form}, for the case of a single band.

At higher orders, the semiclassical conductivities are essentially the same, comprised of a ``Drude-like'' and ``Berry-curvature''-like term:
\begin{align}
	\label{eq:second_order_conductivity_semiclassical}
	\sigma^{\mu \alpha \beta} 
	\ &=\ e^3 \ibz f_\text{FD} \left( 
	\frac{\partial^\beta \partial^\alpha v^\mu}{(2\widetilde{\omega})\widetilde{\omega}}
	- i \frac{\partial^\beta F^{\mu\alpha}}{\widetilde{\omega}} \right)\\
	\label{eq:third_order_conductivity_semiclassical}
	\sigma^{\mu\alpha \beta \gamma}
	\ &=\ e^4 \ibz f_\text{FD}\left(
	i \frac{\partial^\gamma \partial^\beta \partial^\alpha v^\mu}{(3\widetilde{\omega})(2\widetilde{\omega})\widetilde{\omega}}
	+ \frac{\partial^\gamma \partial^\beta F^{\mu\alpha}}{(2\widetilde{\omega}) \widetilde{\omega}} \right)
\end{align}
where $\widetilde{\omega}$ should be read as $\omega + i\gamma$. 

We will show that these equations reproduce the leading order divergences at $\omega \to 0$ for the quantum calculations of the second- and third-order conductivities. However, the numerous other terms in the quantum formulas are not captured here due to their essential interband nature. It would be interesting to examine a modified semiclassical picture involving interband corrections, which should be able to reproduce more of the full response. 

Under time-reversal symmetry, $\v{\nabla}_{\v{k}}$, $\v{v}$, $F$, and $\v{k}$ all change sign, so at second order only the derivative of the Berry curvature survives, while at third order only the velocity term remains. One can extrapolate the pattern in \eqref{eq:third_order_conductivity_semiclassical} to all orders in semiclassics.

\subsection{Length Gauge}

Let us now derive the semiclassical limit starting in the length gauge formulation, \eqref{eq:length_gauge}. We adopt the standard density-matrix approach pioneered by Sipe and Shkrebtii\cite{sipe2000second}, defining the single-particle reduced density matrix
\begin{equation}
\rho_{\v{k}ab}(t) =\braket{c^\dagger_{\v{k}a}(t) c_{\v{k}b}(t)}.
\label{eq:reduced_density_matrix}
\end{equation}
Then the current is given by $J^\mu(t) = e \Tr\left[ \widehat{v}^\mu \rho(t) \right]$, where the trace is taken over the single-particle Hilbert space, i.e. it stands for the integral of the Brillouin zone and sum over bands.

The time-dependence of the density matrix in the interaction picture is given by the Schwinger-Tomonaga Equation
\begin{equation}
	i \frac{d}{dt} \widehat{\rho}_I(t) = \left[ \widehat{H}_{E,I}(t), \widehat{\rho}_I(t) \right],
	\label{eq:schwinger_tomonaga}
\end{equation}
where the subscript $I$ indicates the interaction picture: $\widehat{\mathcal{O}}_I(t) = U(t)^\dagger \widehat{\mathcal{O}} U(t)$ for $U(t) = e^{-i t \widehat{H}_0}$. We can solve \eqref{eq:schwinger_tomonaga} within the framework of perturbation theory by employing Dyson series. Expand $\widehat{\rho} = \sum_{K} \widehat{\rho}^{(K)}$ as a power series in power of the electric field. We can then integrate \eqref{eq:schwinger_tomonaga} to find an order-by-order solution
\begin{equation}
	\widehat{\rho}^{(K+1)}(t) = -i \int_{-\infty}^t d\tau \; \left[ \widehat{H}_{E,I}(\tau), \widehat{\rho}^{(K)}_I(\tau) \right].
	\label{eq:order_by_order_solution_length_gauge}
\end{equation}
This also requires an initial condition $\widehat{\rho}^{(0)} = \delta_{ab} f_a$, taken to be the Fermi-Dirac distribution. We can now write a computable expression for the current. At $n$th order, the current can be written as a nested commutator
\begin{align}
	\label{eq:J_length_gauge}
	J^{\mu}(t) &= e\prod_{k=1}^n \int_{-\infty}^{\tau_{k-1}} d\tau_{k} (ie E^{\alpha_k})\\
	\nonumber
	\ &\hspace{2em}\times\ \Tr\Big\{ \left[ \cdots [\widehat{v}^\mu, \widehat{r}^{\alpha_1}(\tau_1)] \cdots, \widehat{r}^{\alpha_n}(\tau_n) \right]  \widehat{\rho}_0 \Big\}
\end{align}
where $\tau_{-1} \equiv t$. 
Rearranging commutators and Fourier transforming, the $n$th nonlinear conductivity can then be written
\begin{align}
	\label{eq:length_gauge_conductivities}
	&\sigma^{\mu\alpha_1\dots\alpha_n}(\omega;\omega_1,\dots,\omega_n)
	= \frac{1}{n!} \mathcal{S}_n e \prod_{k=1}^n \int_{-\infty}^{\tau_{k-1}} d\tau_k e^{-i \omega_k \tau_k}\\
	\nonumber
	&\hspace{2em}\times (ie) \Tr\Big\{ \widehat{\rho}_0 
	\left[ \widehat{r}^{\alpha_n}(\tau_n), \cdots, [\widehat{r}^{\alpha_1}(\tau_1), \widehat{v}^\mu] \cdots \right]
\Big\}
\end{align}
where $S_n$ symmetrizes over all incoming frequencies $\st{(\alpha_k, \omega_k)\; : \; 1\le k \le n}$.

Our task is now to evaluate this commutator at leading order in $\omega$. This is done most expediently by using the relation between the position operator and the covariant derivative $\widehat{\v{r}} = i \widehat{\mathcal{\v{D}}}$, which is described in Appendix \ref{app:connections}. The form of the commutators in \eqref{eq:length_gauge_conductivities} is almost the same as the covariant derivative repeatedly acting on the velocity operator---but we must account for the time dependence. The time-evolved operator $\widehat{r}(t) = U(t)^\dagger\, \widehat{r} \, U(t)$ is easily computed by noting that, in the energy basis, $U(t)_{ab} = e^{-i \varepsilon_{\v{k}a}(t)} \delta_{ab}$ is a one-parameter family of gauge transformations. Equation \eqref{eq:U(N)_gauge_transformation} implies
\begin{equation}
	i \widehat{\v{D}}(t) = i\v{\nabla} + \v{A}'(t), \v{A}'(t) = e^{iH_0t} \v{A} e^{-iH_0t} + t \nabla H_0
\end{equation}
where $(\v{\nabla}  \widehat{H}_0)_{ab} = \delta_{ab} \v{\nabla}_{\v{k}} \varepsilon_a(\v{k})$ is the regular gradient of the matrix elements. In components, this implies the identity
\begin{align}
	\label{eq:time_evolution_commutator_action}
	[\widehat{r}^\alpha(\tau), \mathcal{O}]_{ab} \ &=\ \left( i \partial^\alpha + \tau \Delta_{ab}^\alpha \right) \mathcal{O}_{ab}\\
	\nonumber
	\ & \hspace{2em}+ \sum_{c} e^{i \varepsilon_{ac} t} A^\alpha_{ac} \mathcal{O}_{cb} - \mathcal{O}_{ac} A^\alpha_{cb} e^{i \varepsilon_{cb} t},
\end{align}
where we have defined $\Delta_{ab}^\alpha \equiv h_{aa}^\alpha - h_{bb}^\alpha = \partial^\alpha \varepsilon_{ab}$.  

Employing \eqref{eq:time_evolution_commutator_action} with $\widehat{\mathcal{O}} = \widehat{v}^\mu$,
\begin{align}
	&\sigma^{\mu\alpha}(\omega;\omega) = \frac{ie^2}{\hbar} \sum_{a,b} \ibz f_a \int_{-\infty}^0 e^{-i \omega \tau}\\
	\nonumber
	&\hspace{2em} \times \left( i \partial^\alpha v^\mu + e^{i \varepsilon_{ab} \tau} A^\alpha_{ab} v^\mu_{ba} - e^{i \varepsilon_{ba} \tau} v^\mu_{ab} A^\alpha_{ba} \right).
\end{align}
As is customary, when performing the time integral, a phenomenological relaxation rate $\omega \to \omega + i \gamma$ is added so that $\int_{-\infty}^0 d\tau \; e^{i(\zeta-\omega-i\gamma)\tau} = \frac{i}{\omega+i\gamma - \zeta}$. Therefore the linear conductivity is
\begin{equation}
	\sigma^{\mu\alpha}(\omega;\omega) = \frac{ie^2}{\hbar} \sum_{a \neq b} \ibz f_a \frac{\partial^\alpha v^\mu}{\omega + i \gamma} + f_{ab} \frac{A^\alpha_{ab} v^\mu_{ba}}{\varepsilon_{ab} - \omega - i \gamma}.
	\label{eq:linear_conductivity_length_gauge}
\end{equation}
The first term reproduce the Drude formula, and the second term is almost $[A^\alpha,A^\mu]$. This is equivalent to the expression derived in the velocity gauge, Equation \eqref{eq:conductivity_no_spurious_divergence}. Under the limit $\omega \ll \varepsilon_{ab}$, one arrives at \eqref{eq:conductivity_drude_form}, which matches the semiclassical result \eqref{eq:semi_classical_linear_conductivity} at linear order in $\v{E}$.

At nonlinear order, one must evaluate further nested commutators. Since we are only interested in the $\omega \to 0$ limit, we will limit ourselves to the leading order terms. However, this procedure can be easily continued to give expressions for the general conductivity tensors in the length gauge; such a calculation is carried out in \cite{ventura2017gauge}. At second order, we consider the expression $[\widehat{r}^\beta(\tau_2), [\widehat{r}^\alpha(\tau_1), \widehat{v}^\mu]]_{aa}$. Expanding, 
\begin{align}
	[\widehat{r}^\beta(\tau_2), [\widehat{r}^\alpha(\tau_1), \widehat{v}^\mu]]_{aa}
	\ &=\ i \partial^\beta i \partial^\alpha v^\mu_{aa}\\
	\nonumber
	\ &\hspace{1em} + i \partial^\beta \left[ A^\alpha(\tau_1), v^\mu \right]_{aa}\\
	\nonumber
	\ &\hspace{1em} + \left[ A^\beta(\tau_2), (r^\alpha(\tau_1) v^\mu) \right]_{aa}\\
	\nonumber
	\ &\hspace{1em} + \left[ A^\beta(\tau_2), [A^\alpha(\tau_1), v^\mu] \right]_{aa}
\end{align}
where $A^\alpha_{ab}(\tau) \equiv e^{i\varepsilon_{ab} \tau} A^\alpha_{ab}$ is the time-evolved operator. Each factor of $e^{i \varepsilon_{ab} \tau}$ is Fourier transformed to a denominator of the form $\frac{1}{\omega -\varepsilon_{ab}}$. However, the number of such exponential factors is different in each term. The first term has none, the second term has one, and the latter terms generically have two. The Fourier transform of the first two terms is therefore
\begin{equation}
	\label{eq:second_order_semiclassical_terms}
	\frac{\partial^\beta \partial^\alpha v^\mu_{aa}}{\omega \omega_{2}} + \sum_{b\neq a} \frac{-i \partial^\beta}{\omega_2}\left[ \frac{A^\alpha_{ab} v^\mu_{ba}}{\omega-\varepsilon_{ba}} - \frac{v^\mu_{ab} A^\alpha_{ba}}{\omega-\varepsilon_{ab}} \right].
\end{equation}
After Fourier transforming the third and fourth terms, either there are factors $\frac{1}{\varepsilon_{cd} -\omega}$, which are $O(\omega^0)$ and hence subleading or, when $c=d$ there are poles $\frac{1}{\omega}$, which cancel out due to the commutator in the $\omega \to 0$ limit. Hence only the terms \eqref{eq:second_order_semiclassical_terms} survive in the semiclassical limit, so
\begin{align}
	&\lim_{\omega, \omega_i \to 0} \sigma^{\mu\alpha\beta}(\omega, \omega_1, \omega_2)= \frac{-e^3}{\hbar^2} \sum_{a,b} \ibz
\frac{f_a \partial^\beta \partial^\alpha v^\mu_{aa}}{\omega \omega_{2}}\\
\nonumber
&\hspace{4em}+ f_{a}\frac{-i \partial^\beta}{\omega_2}\left[ \frac{A^\alpha_{ab} v^\mu_{ba}}{\omega-\varepsilon_{ba}} - \frac{v^\mu_{ab} A^\alpha_{ba}}{\omega-\varepsilon_{ab}} \right] + O(\omega^0)
\end{align}
or
\begin{align}
	&\lim_{\omega,\omega_i \to 0} \sigma^{\mu\alpha\beta}(\omega, \omega_1, \omega_2)= \frac{-e^3}{\hbar^2} \sum_{a} \ibz \frac{f_a \partial^\beta \partial^\alpha v_{aa}^\mu}{\omega \omega_2}\\
	\nonumber
	& \hspace{4em}+ f_a \frac{-i \partial^\beta}{\omega_2} \mathcal{F}^{\alpha\mu}_{aa} + O(\omega^0).
	\label{eq:second_order_semiclassical_limit_length_gauge}
\end{align}
The two terms are clearly just $\frac{ie}{\hbar}\frac{\partial^\beta}{\omega_2}$ acting on the first order expression --- exactly in line with the semiclassical prediction \eqref{eq:second_order_conductivity_semiclassical}. The first term is the derivative of the Drude weight, while the second is the Berry curvature dipole, which was studied in semiclassics \cite{orensteinmoore,sodemannfu} and with a Floquet formalism \cite{morimoto2016semiclassical}. As mentioned above, this is the only term that survives in the presence of time-reversal symmetry.

At third order, one must consider
\begin{align}
	&\left[ \widehat{r}^\gamma(\tau_3), [\widehat{r}^\beta(\tau_2), [\widehat{r}^\alpha(\tau_1),\widehat{v}^\mu]] \right]_{aa}\\
	\nonumber
	&\hspace{2em}= i\partial^\gamma i\partial^\beta i\partial^\alpha v^\mu_{aa} + i\partial^\gamma i\partial^\beta [A^\alpha(\tau), v^\mu]_{aa} + \cdots.
\end{align}
Due to the same logic that applied at second order, only these first few times survive at lowest order in $\omega$. Hence
\begin{align}
	&\sigma^{\mu\alpha\beta\gamma}(\omega;\omega_1,\omega_2,\omega_3) = \frac{e^4}{\hbar^3} \sum_{a,b} \ibz f_a  \frac{\partial^\gamma \partial^\beta \partial^\alpha v^\mu_{aa}}{\omega \omega_{23} \omega_3}\\
	\nonumber
	\ & \hspace{3em}-i f_a \frac{\partial^\gamma \partial^\beta}{\omega_{23} \omega_{3}}\left[ 
			\frac{A^\alpha_{ab} v^\mu_{ba}}{\omega-\varepsilon_{ba}} - \frac{v^\mu_{ab} A^\alpha_{ba}}{\omega-\varepsilon_{ab}} 
	\right]
	+ O(\omega^{-1})
\end{align}
so
\begin{align}
	\label{eq:third_order_conductivity_semiclassical_limit_length}
	&\lim_{\omega,\omega_i\to 0} \sigma^{\mu\alpha\beta\gamma}(\omega;\omega_1,\omega_2,\omega_3) =\\
	\nonumber
	&\frac{e^4}{\hbar^3} \sum_{a,b} \ibz f_a  \frac{\partial^\gamma \partial^\beta \partial^\alpha v^\mu_{aa}}{ \omega_{23} \omega_3\omega}
	-i f_a \frac{\partial^\gamma \partial^\beta}{\omega_{23} \omega_{3}} \mathcal{F}^{\alpha\mu}_{aa}+ O(\omega^{-1})
\end{align}
Here the first term is the third derivative of the Drude weight while the second is the Berry curvature quadrapole. Of course, this matches the semiclassical result \eqref{eq:third_order_conductivity_semiclassical}.

\subsection{Symmetry Considerations}
\label{subsec:symmetry_considerations}

This subsection considers the effect of symmetry on the two terms of the semiclassical third order response. We focus on the effect of inversion $\mathcal{I}$, time-reveral $\mathcal{T}$, and reflection in the $b$ direction $\mathcal{R}^b$. The semiclassical response involves the group velocity $v^m$, Berry curvature $\mathcal{F}^{\alpha\mu}$, and $k$-derivatives thereof, so we start by looking at their transformations under symmetry. Their transformation laws can be deduced from the fact that $v^\mu$ and $\partial^\alpha$ are vectors, while the Berry curvature behaves as a psuedovector defined by $\mathcal{F}^{\beta} \equiv \varepsilon_{\beta\alpha\mu} \mathcal{F}^{\alpha\mu}/2$. 

The effect of inversion $\mathcal{I}$ is 
\begin{align}
v^\mu &\to - v^\mu \\
\mathcal{F}^{\beta} & \to \mathcal{F}^{\beta} \\
\partial^\alpha &\to -\partial^\alpha. 
\end{align}
Applying time reversal $\mathcal{T}$ gives
\begin{align}
v^\mu &\to - v^\mu \\
\mathcal{F}^{\beta} & \to - \mathcal{F}^{\beta} \\
\partial^\alpha &\to -\partial^\alpha. 
\end{align}
Lastly, the reflection $\mathcal{R}^b$ leads to
\begin{align}
v^\mu &\to (-1)^{\delta_{b \mu}} v^\mu \\
\mathcal{F}^{\beta} & \to - (-1)^{\delta_{b \beta}} \mathcal{F}^{\beta} \\
\partial^\alpha &\to (-1)^{\delta_{b \alpha}} \partial^\alpha. 
\end{align}
These constraints indicate that the group velocity term and the Berry curvature term (the first and the second terms in Eq. \eqref{eq:third_order_conductivity_semiclassical_limit_length}, respectively) are both even under $\mathcal{I}$, and even and odd under $\mathcal{T}$, respectively. 
Under $\mathcal{R}^b$, either both terms are even or both terms are odd, depending on the component of nonlinear conductivity and the direction of mirror plane.
For example, in $\sigma^{zxxx}$, both are odd under $\mathcal{R}^z$, and both are even under $\mathcal{R}^y$.

An interesting question is when the group velocity term vanishes, whereupon the Berry curvature contribution dominates, if it is non-zero. First, this requires $\mathcal{T}$ breaking. Next we need a symmetry such that the group velocity term is odd and the Berry curvature is even. Such a symmetry is obtained by combining $\mathcal{R}^b$ in which both terms are odd and $\mathcal{T}$.
For example, $\sigma^{zxxx}$ has a nonzero contribution only from the Berry curvature term when $\mathcal{T R}^z$ symmetry is preserved and both $\mathcal{T}$ and $\mathcal{R}^z$ symmetries are broken.
This situation can be realized in materials with antiferomagnetic order in $z$ direction.
Since $\sigma^{zxxx}$ is measured as intensity dependent Hall conductivity or intensity dependent transmission of circular polarized light,
measuring these quantities in suitable antiferromagnetic materials will allow us to access the Berry curvature effect in third order responses.
This Berry curvature effect might be measured in the magnetic Weyl semimetal $\text{Mn}_3\text{Sn}$\cite{nakatsuji2015large,kiyohara2016giant,nayak2016large} since it breaks $\mathcal{T}$ and some candidate AFM structures break $\mathcal{R}^z$ while preserving $\mathcal{T R}^z$ \cite{Suzuki17}.

\subsection{Length versus Velocity Gauges}

Overall, we have shown that the semiclassical limit is straightforwardly accomplished in the length gauge and matches the answer from the simple Boltzmann equation approach. A few comments on the relation between the length and velocity gauge are in order. It was shown in \cite{passos2017nonlinear} that on can convert between the two gauges with the time-dependent unitary transformation $S(t) = e^{-\frac{e}{\hbar}\v{A}(t) \cdot \v{D}}$. The equivalence of expectations of any physical observable $\mathcal{O}$ in the two gauges leads to sum rules of the form\cite{ventura2017gauge}
\begin{equation}
	\ibz \Tr\Big\{ D^{\alpha_1} \cdots D^{\alpha_n} \left[\mathcal{O} \widehat{\rho}_{\v{k}}(t)\right] \Big\} = 0
\end{equation}
where $\rho_{\v{k}}(t)$ is the single-particle density matrix defined above. Expanding this with $\mathcal{D} = i \nabla + \mathcal{A}$ leads to the sum rules of Aversa and Sipe\cite{aversa1995nonlinear}. In particular, one can use $\mathcal{O} = \widehat{v}^m$ to convert from velocity to length gauge at order $n$. This will eliminate terms like $h^{\mu\alpha}$ in favor of $\partial^\alpha v^\mu + \cdots$. However, this algebra is quite involved in practice, so it is usually better to choose the correct gauge from the outset rather than painstakingly changing gauge after writing the answer to a computation.

\section{Numerical Example}
\label{sec:numerical_example}

This section applies the techniques developed in this paper to a model of Weyl semimetals. Numerical calculations of nonlinear optics are usually done within the frameworks of either tight-binding models or Density Functional Theory (DFT). Tight-binding models are usually simple enough to perform analytical calculations and, when chosen wisely, will reproduce the main qualitative features of a material, such as the frequencies of resonances. For more quantitative predictions in specific materials, DFT is the favored technique. Velocity gauge formulas are particularly well-suited for tight-binding models, where operators such as $h^{\mu\alpha}$ may be computed analytically. We will therefore present the example of a simple tight-binding model of a Weyl semimetal where the leading contribution is topological in origin.

The primary feature of Weyl semimetals are their paired Weyl- and anti-Weyl cones, whose linear dispersion acts as a sources and sinks of Berry curvature. As mentioned in the introduction, a wide variety of linear and second order optical responses have been studied in Weyl semimetals, many with a topological origin. Here we study the third-order response $\sigma^{zxxx}$, for which the leading contribution comes from the (topological) Berry curvature. To our knowledge, this is the first prediction for a third-order response in Weyl semimetals.

Consider the following two-band Hamiltonian for a Weyl semimetal with a Wilson mass:
\begin{equation}
	H(\v{k}) = d_0 I + \v{d}(\v{k})\cdot \v{\sigma}	
	\label{eq:weyl_tight_binding_model}
\end{equation}
where $\v{\sigma} = \st{\sigma_x,\sigma_y,\sigma_z}$ is the vector of Pauli matrices, $\v{d} = \st{d_x, d_y, d_z}$, and
\begin{align*}
	d_0(\v{k}) \ &=\ t \sin ak_y\\
	d_x(\v{k}) \ &=\ \sin a k_x\\
	d_y(\v{k}) \ &=\ \sin a k_y\\
	d_z(\v{k}) \ &=\ \cos a k_z + m\left( 2 - \cos a k_x - \cos a k_y \right),
\end{align*}
where $a$ is the lattice spacing. Generically, this model supports four Weyl--anti-Weyl pairs, but we gap out three of them by adding Wilson mass term where we set $m=1$ \cite{Vafek14}. The remaining Weyl nodes are at $\v{k} = (0,0, \pm \pi/2)$. The parameter $t$ controls the tilting of the Weyl nodes.

In Section \ref{subsec:symmetry_considerations}, we showed that materials where time-reversal symmetry $\mathcal{T}$ and a mirror symmetry $\mathcal{R}^z$ are broken, but their product $\mathcal{T}\mathcal{R}^z$ is preserved, then the leading order contribution at third order in the $\omega \to 0$ limit is

\begin{align}
	\label{eq:limit_third_order_leading_Berry_limit}
	&\sigma^{\mu\alpha\beta\gamma}(\omega;\omega_1, \omega_2, \omega_3)\\
	\nonumber
	& = -i \frac{e^4}{\hbar^3} \sum_{a} \ibz  \frac{f_a \partial^\gamma \partial^\beta \mathcal{F}^{\alpha\mu}_{aa}}{\left( \omega_3 + i \gamma \right) \left( \omega_2 + \omega_3 + 2i\gamma \right)} + O(\omega^{-1}).
\end{align}
Whenever the tilting parameters $t$ is nonzero, the model satisfies these considerations and thus we expect a topological leading response in the off-diagonal component of the third-harmonic response $\sigma^{zxxx}(3\omega;\omega,\omega,\omega)$. (Here $z$ is the direction of the emitted light.) The tilting is selected to be in the $y$-direction so that both nodes are tilted the same way, making the resonances symmetry-allowed.
\begin{figure}
	\centering
	\includegraphics[width=\linewidth]{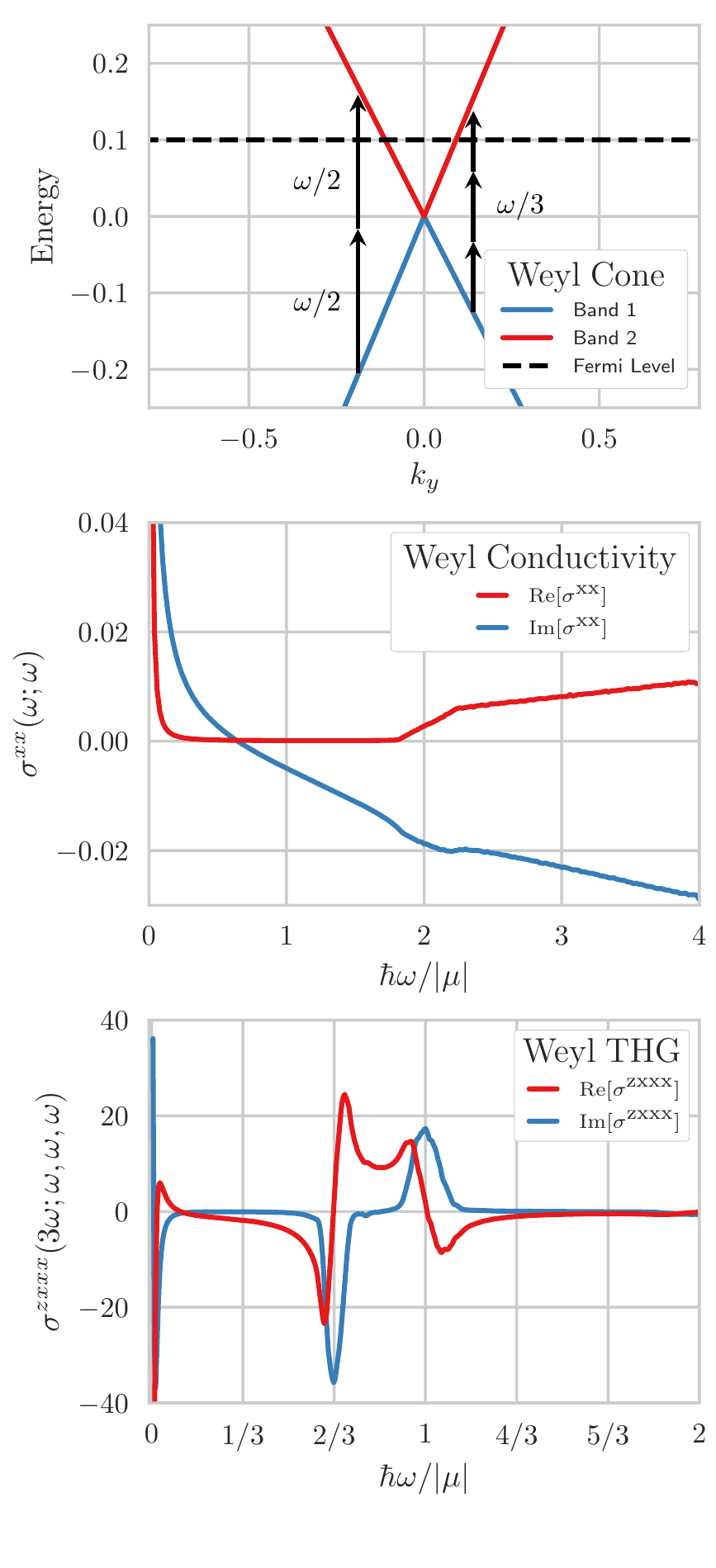}
	\caption{\label{fig:weyl_spectra}
		\textit{(Top)} Band structure of Equation \eqref{eq:weyl_tight_binding_model} along the $k_x = 0$, $k_z = \pi/2$ plane. A (tilted) Weyl node is visible at $k_y = 0$. The dispersion is approximately linear for $\n{\varepsilon} \le 8 \mu$.
		\textit{(Middle)} Linear conductivity in the $x$-direction. The Drude peak is visible at low frequencies, the conductivity increases linearly for $\omega \gg 2 \mu$.
		\textit{(Bottom)} Off-diagonal component of the third harmonic response. The $O(\omega^{-2})$ divergence due to the Berry curvature is visible at low frequencies, and wide resonances are visible at $\omega \sim{} 2\mu/3$ and $\omega \sim{} \mu$.  The parameters used for all data are $\mu = 0.1$, $\gamma = 0.001$, $t = 0.1$, $m=a=1$.} 
\end{figure}

We compute the response via numerically integrating Equation \eqref{eq:third_harmonic_generation} for the third-harmonic on a mesh of $k$-points until convergence is achieved. This involves evaluating the band energies, wavefunctions, and higher derivatives of the Hamiltonian, which may all be computed analytically. As usual for a two-band model, the energies and wavefunctions are, up to normalization,
\begin{equation}
	\varepsilon_{\pm}(\v{k}) = d_0 \pm \n{\v{d}}, \quad
	\ket{u_{\pm}} =
	\begin{pmatrix}
		\frac{d_3 \pm \n{\v{d}}}{d_1 + i d_2}\\
		1
	\end{pmatrix}.
	\label{eq:eigenvalue_and_vectors_Weyl}
\end{equation}
The velocity operators are easily found by differentiating:
\begin{align}
	h^x(\v{k}) \ &= \ t a \cos a k_y I + a \cos a k_x \sigma_x + m a \sin a k_x \sigma_z,\\
	h^y(\v{k}) \ &=\ a \cos a k_y \sigma_y + m a \sin a k_y \sigma_z,\\
	h^z(\v{k}) \ &=\ - a \sin k_z \sigma_z.
\end{align}
Higher derivatives are similarly straightforward. For example,
\begin{align}
	h^{xx}(\v{k}) \ &=\ -t a^2 \sin a k_y I - a^2 \sin a k_x \sigma_x\\
	\ &\hspace{2em} + m a^2 \cos a k_x \sigma_z,\\
	h^{xxx}(\v{k}) \ &=\ - a^2 h^{x}(\v{k}),\\
	h^{xxxx}(\v{k}) \ &=\ - a^2 h^{xx}(\v{k}).
\end{align}
Figure \ref{fig:weyl_spectra} shows the well-understood linear reponse of the material and the third-harmonic response $\sigma^{zxxx}$.

The linear response of Weyl semimetals is shown in the middle panel of FIG \ref{fig:weyl_spectra}. This response is already well-understood \cite{Hosur}. The conductivity exhibits a typical Drude peak at low frequencies, and $\re[\sigma^{xx}] \sim{} \omega$ at higher frequencies. This is due to the interband resonance that becomes possible once the frequency exceeeds twice the chemical potential. Due to the tilting, the Fermi surface with $\mu = 0.1$ is an ellipse with a smaller bandgap on one side than another. This causes the linear conducitivity to onset in the range $2\mu - t \le \omega \le 2\mu + t$.

The third harmonic response is shown in the bottom panel of FIG. \ref{fig:weyl_spectra}, and displays several features of interest. At low frequency, where $\omega \sim \gamma$, we observe the predicted divergence \eqref{eq:limit_third_order_leading_Berry_limit}, in accordance with the semiclassical considerations of Section \ref{sec:semiclassics}. The divergence is visible in both the real and imaginary parts due to the phenomenological broadening. Resonances are visible near $\omega \sim{} \mu$ and $\omega \sim{} 2\mu/3$, due to two- and three-photon processes respectively. The top panel of FIG. \ref{fig:weyl_spectra} indicates these processes schematically.

One can see that the two-photon process becomes resonant around $\omega \sim{} 2\mu/3 - t$, corresponding to the side of the Weyl cone with the smaller bandgap, and continues up to $\omega \sim{} 2\mu/3 + t$, when the resonance is on the other side of the cone. This causes a peculiar linear increase in $\re{\sigma^{zxxx}}$ in the range $2\mu/3 -t \le \omega \le 2\mu/3 +t$, and similar considerations apply to the third-order response in the range $\mu -t \le \omega \le \mu + t$. As the tilting is increased, the range of this linear regime grows. This not too surprising, since a similar linear onset due to the tilt is present at first order. To our knowledge, this is the first prediction of a third-order response in Weyl semimetals. Again, the key topological feature is the divergence at low frequency, which is proportional to the quadrupole moment of the Berry curvature.

One should note that, although we have focused on the third-harmonic contribution here, the equations for the third-order response from Section \ref{sec:third_order} are generic. One can just as easily evaluate the self-focusing correction, totally off-diagonal components such as $\sigma^{zyxz}$, or other effects such as the AC Kerr effect. Similarly, any other tight-binding model can be used instead of \eqref{eq:weyl_tight_binding_model}. The only restriction is that it must be defined on the entire Brillouin zone, so that the equivalence with length-gauge is maintained.

Let us comment briefly on the use of DFT. The optical response formulas in previous sections require the matrix elements of derivatives of the Hamiltonian operator \eqref{eq:higher_derivatives_of_H}. Using the covariant derivative \eqref{eq:operator_derivative_matrix_elements}, these can be written in terms of the matrix elements of the velocity operator and the Berry connection. There are well-established techniques for calculating linear responses within DFT\cite{rohlfing2000electron}---which already involves computing the Berry connection and matrix elements of the velocity operators---and frequently achieves predictions within $1\%-10\%$ of experimental values. Naively, this is somewhat surprising, as DFT does not necessarily give good wavefunctions, but only energies. Nevertheless, tools such as the ``GW'' approximation or the use of specific functionals permit accurate determination of the wavefunctions in many cases. With sufficiently fine $k$-space meshes, one can in principle converge the numerical derivatives required and make accurate predictions for non-linear optical responses within DFT\cite{rangel2017large}. Another option is to use the technique of ``Wannierization'' to produce accurate tight-binding models by Fourier transforming Wannier functions derived \textit{ab initio} \cite{ibanez2018ab}. In sum, the nonlinear optical responses presented here may, in principle, be accurately computed within DFT.

\section{Discussion and Conclusions}
\label{sec:conclusions}

This work has elucidated a diagrammatic approach to nonlinear optical responses and applied it to predict the third order optical response of Weyl semimetals. In this final section we will reiterate the main results of the paper and discuss the choice of gauge.

As mentioned in the introduction, the choice of the length or velocity gauge in optical response calculations is a longstanding issue. The modern definitions of the gauges---which depend crucially on the Berry connection---permit the use of either gauge to compute optical responses. Therefore one is now free to choose the best gauge for the problem at hand. The computations in this work suggest a few rules of thumb for when each gauge should be applied. Equations in the velocity gauge have a natural separation between matrix elements and resonances, and contain only simple poles, making them preferable whenever it is necessary to separately examine one-, two-, and three-photon resonances. Since the only matrix elements that appear are derivatives of the Hamiltonian, the velocity gauge is particularly well-suited for tight-binding calculations. However, in the $\omega \to 0$ limit, the velocity gauge suffers from (cancelling) apparent divergences. Hence, for analytical work in this limit the length gauge is often preferable.

Let us comment on why our diagrammatic approach necessarily employs the velocity gauge. The key issue is the presence of the position operator $\widehat{\v{r}}$, which acts on \textit{all operators to the right} by differentiation. The vertices needed in the length gauge become complicated quite quickly, as they involve not only the position and velocity operators, but objects such as the derivatives of the position operator and a resonance; virtually every term uses its own, unique, vertex. A naive diagrammatic approach to nonlinear response in the length gauge is therefore impractical. One should note that, historically, diagrammatic methods have indeed employed the length gauge \cite{Ward65}. However, these techniques do not account for the Berry connection, but only the fully interband parts of the position operator. In any material with non-vanishing Berry connection, these old-style diagrams will miss important contributions to nonlinear responses, including some resonances. 

The diagrammatic method of this work provides an efficient computational framework to calculate nonlinear responses in the velocity gauge. The results are general for any component and frequency, without unphysical divergences. We have provided convenient formulae for the general second and third order responses, as well as the particular cases of second harmonic, shift current, third harmonic and self-focusing. To interpret these equations, we examined the semiclassical limit and linked it to the length gauge. On a technical level, the method of this work should often be the shortest way to compute nonlinear optical responses. 

The expressions for nonlinear optical responses given here are equivalent to those previously given in the literature in all cases we are aware of (so long as the correct definitions for the length and velocity gauges are employed). We have checked that our formalism explicitly reproduces the results of Refs \cite{sipe2000second,orensteinmoore,sodemannfu,Chan16,PhysRevB.93.201202, PhysRevLett.117.216601}, as well as the equivalence of our equations for the first-order conductivity, shift current, and second-harmonic generation with those present in the literature. This is exactly what is expected. After all, one can recover many other schemes for computing non-linear responses as limits of ours, including (i) Boltzmann/semiclassical transport theory, (ii) quantum mechanical perturbation theory in the length or velocity gauge, (iii) Floquet formalism. Recent work \cite{joao2018non} develops a diagrammatic expansion for non-linear optical responses in the Keldysh formalism which reduces to our formalism when the applied electric fields are periodic in time (i.e. plane-waves). We expect, however, that our results hold for general wavepackets $\v{E}(t)$ so long as the duration of the wavepacket and measurement are much less than the timescale associated with dissipation.

Optical responses are most useful when connected to experiment. To this end, we have predicted the third harmonic response of a Weyl semimetal. At small frequencies, the third harmonic response is dominated by a divergent term due to the quadrupole of the Berry curvature, and hence of topological origin. There are also large resonant contributions from both two- and three-photon processes, with a peculiar linear character.

The results of this work can be expanded in both technical and practical directions. Technically, the diagrammatic formalism enables interacting electrons to be treated on the same level as free ones; we are currently expanding these results to the case of Fermi liquids and possibly even magnetic fields. On a practical level, third order responses are somewhat understudied at present, despite being present in most materials and technologically important. The formulae and techniques of this work should enable or simplify prediction of the third order optical response in a wide variety of materials.

\begin{acknowledgements}

	The authors thank Adolfo Grushin, Daniel Passos, and Jonah Haber for useful discussions. This work was primarily funded by the U.S. Department of Energy, Office of Science, Office of Basic Energy Sciences, Materials Sciences and Engineering Division under Contract No. DE-AC02-05-CH11231 (Quantum materials program KC2202).  T.M was supported by the Gordon and Betty Moore Foundation’s EPiQS Initiative Theory Center Grant to UC Berkeley. J.O. received support from the Gordon and Betty Moore Foundation’s EPiQS Initiative through Grant GBMF4537. D.E.P. received support from the NSF GRFP, DGE 1752814.
\end{acknowledgements}

\appendix

\section{The Position Operator and The Berry Connection}
\label{app:connections}

This Appendix discusses the position operator and its close relation to the Berry connection, giving some mathematical details thereof.

\subsection{The Position Operator as Covariant Derivative}

Suppose we have a crystal with a finite number of bands, $N$, which are all close to the Fermi level and separated from all other bands by a large energy gap. We can then consider those $N$ bands as an effective model for the material. What form does the single-particle position operator take in this situation? The correct answer to this question was known at least as early as 1962, where it is discussed in the classic paper of Blount\cite{blount1962formalisms}.  Morally, just as derivatives and polynomials are exchanged by Fourier transforms, the real-space position operator $\widehat{\v{r}}$ should become a $\v{k}$-derivative. We briefly recall Blount's derivation, adapted to modern notation. 

Any wavefunction $\ket{f}$ can be written in terms of the Bloch functions $\psi_{\v{k}a}$ as
\begin{equation}
	\braket{\v{r}|f} = f(\v{r}) = \sum_a \ibz \psi_{\v{k}a} f_a(\v{k})
\end{equation}
Then
\begin{align*}
	\braket{\v{r}|\widehat{\v{r}}|f} \ &=\ \sum_{a} \ibz\; \psi_{\v{k}a}(\v{r}) \v{r} f_a(\v{k})\\
	\ &=\ \sum_{a} \ibz \;  \left[-i\partial_{\v{k}} \left( e^{i \v{k}\cdot \v{r}} \right)\right] u_{\v{k}a}(\v{r}) f_a(\v{k})
\end{align*}
Integrating by parts (the surface term vanishes because the Brillouin zone is a closed manifold)
\begin{align*}
		\braket{\v{r}|\widehat{\v{r}}|f}
		\ &=\ \sum_a \ibz \; e^{i \v{k}\cdot \v{r}} \left[  i \partial_{\v{k}} u_{\v{k}a}(\v{r}) + u_{\v{k} a}(\v{r}) i \partial_{\v{k}} f\right]\\
		\ &=\ \sum_{a,b} \ibz \;  \psi_{\v{k}b}(\v{r})\left[ \delta_{ab} i\partial_{\v{k}} + u_{\v{k}b} i \partial_{\v{k}} u_{\v{k}a} \right] f_a.
\end{align*}
We can therefore identify 
\begin{equation}
	\widehat{\v{r}} = i \widehat{\v{\mathcal{D}}} = i \left[ \v{\nabla}_{\v{k}} - i \v{\mathcal{A}} \right]
	\label{eq:covariant_derivative_definition}
\end{equation}
where
\begin{equation}
	\v{\mathcal{A}}_{ab} = i \braket{u_{\v{k}a}|\partial_{\v{k}} u_{\v{k}b}}.
	\label{eq:connection_definition}
\end{equation}
To be clear, in \eqref{eq:covariant_derivative_definition}, $\v{\nabla}_{\v{k}} = \delta_{ab} \v{\nabla}_{\v{k}} \delta(\v{k}'-\v{k})$ is the gradient operator which acts on \textit{all matrix elements to the right}. Here we have used the standard notation $\psi_{\v{k}a}$ for the Bloch functions, but nowhere was the fact that they are eigenvectors of the Hamiltonian necessary. Indeed, nothing about the Hamiltonian was needed! The connection we have defined is a generalization of the Berry connection to the case of multiple bands; $\v{\mathcal{D}}$ is a $U(N)$ connection. It depends only on the choice of which $N$ bands are involved and not on any details of the dynamics.

This is particularly clear once we consider a change of basis. Suppose $U$ is a general change of basis, i.e. a $U(N)$ gauge transformation: $\psi^{'}_{\v{k}a'} = U_{a'a}(\v{k}) \psi_{\v{k} a}$ where the $U_{a'a}$'s vary smoothly with $\v{k}$. Gauge transforms act naturally on basis vectors, and therefore act through the dual representation on wavefunctions, which are coefficients. Concretely, $\braket{\v{r}|f} = \sum_a \ibz \psi_{\v{k}a} f_a$ transforms to  
\begin{equation}
	\sum_{a'} \ibz \psi^{'}_{\v{k}a'} f_{a'} = 
	\sum_{a,a'} \ibz \psi_{\v{k}a} U_{aa'}(\v{k}) f_{a'}.
\end{equation}
Hence wavefunctions transform as $f \to U^\dagger f$. We therefore mandate that $\widehat{\v{\mathcal{D}}} f$, which is itself a wavefunction, must transform as 
\begin{equation}
	\widehat{\v{\mathcal{D}}} f \to U^\dagger \widehat{\v{\mathcal{D}}} f = \left( U^\dagger \widehat{\v{\mathcal{D}}} U \right) \left(U^\dagger f \right)
\end{equation}
under a gauge transformation. The action of $\mathcal{\v{D}}$ gives
\begin{equation}
	\left[U^\dagger U \v{\nabla}_{\v{k}} + U^\dagger\left( \v{\nabla}_{\v{k}} U \right) -i U^\dagger \v{A} U \right] \left( U^\dagger f \right).
\end{equation}
Comparing with $\left[ \v{\nabla}_{\v{k}} -i\v{A}' \right] \left( U^\dagger f \right)$ we find 
\begin{equation}
	\v{A}' = U^\dagger \v{A} U + iU^\dagger \v{\nabla}_{\v{k}} U
	\label{eq:U(N)_gauge_transformation}
\end{equation}
This confirms that $\widehat{\v{\mathcal{D}}}$ is a $U(N)$ non-Abelian connection.

Beyond being necessary to define the position operator correctly, this connection allows us to define the $\v{k}$-derivatives of operators. The connection acts on operators naturally via 
\begin{equation}
	\widehat{\mathcal{\v{D}}}[\widehat{\mathcal{O}}] = [\widehat{\mathcal{\v{D}}}, \mathcal{O}].
\end{equation}
This is used extensively in the main text.

\subsection{Generalized Berry Connections}

Let us give a few more mathematical comments. Readers curious for a more formal treatment are recommended to consult Chapter 7 of \cite{bohm2013geometric} or Appendix D of \cite{freed2013twisted}. The normal Berry connection \cite{simon1983holonomy} is a $U(1)$ connection defined for a single band. For our setting of $N$ bands, there are two possible generalizations to consider: a $U(1)^N$ connection or a $U(N)$ connection, the latter of which we have described above. Let us see how each of these arise and what role they play physically.

From the perspective of differential geometry, we are working with an infinite dimensional Hilbert bundle over the Brillouin torus. The exterior derivative $d$ is a provides a (curvature-free) connection on the Hilbert bundle. When we select an $N$-dimensional effective Hilbert space, there is a projection map 
\begin{equation}
	P= \frac{1}{N}\sum_{a=1}^N \ibz \; \ket{u_{\v{k}a}} \bra{u_{\v{k}a}}
\end{equation}
from the Hilbert bundle to the $\C^N$ bundle of interest, and this projection naturally induces a connection on the $\C^N$ bundle which acts on $\C^N$-valued differential forms $\omega$ as\cite{avron1989chern}
\begin{equation}
	D\,\omega = P d \,\omega = (d+i\mathcal{A})\, \omega
	\label{eq:projection_gives_Berry_connection}
\end{equation}

An important, yet subtle, point is that the considerations above do not uniquely define a connection. There is still a residual freedom corresponding to the choice of origin in the (real space) unit cell. This is intimately related to the modern theory of polarization and is carefully considered from a mathematical point of view in \cite{moore2017comment}.

Due to the non-Abelian nature of the $U(N)$ connection, its gauge-invariant quantities are Wilson loops, which cannot be computed directly from the curvature. It would be interesting to compute these and determine if they have any physical meaning or utility. However, it seems unlikely that expressions involving Wilson loops are buried inside nonlinear conductivities. In the special case of degenerate bands, the Wilson loops have been used, for instance, to classify topological parts of Fermi-surface oscillations under magnetic fields \cite{alexandradinata2017topo}.

Now let us identify the second, Abelian, connection. In practice, one virtually always chooses to work in the energy basis with Bloch functions $u_{\v{k}a}$. However, as is well-known, these are only defined up to a phase. The functions $u'_{\v{k}a} = e^{i \theta_a(\v{k})} u_{\v{k}a}$ also satisfy \eqref{eq:single_particle_eigenvalues}. In other words, the choice of the energy basis does not completely fix the gauge, but only up to a change of phase in each band; there is a residual $U(1)^N$ gauge freedom. This allows us to define the second connection, which is Abelian and is denoted by a non-calligraphic letter:
\begin{equation}
	\widehat{\v{D}}(\v{k})_{ab} = \delta_{ab}\left[ \v{\nabla}_{\v{k}} - i \v{A}_{aa} \right]
	\label{eq:multi-band_Berry_connection}
\end{equation}
where $\v{A}$ is the same as above, but this is now \textit{diagonal} in the band indices. Under a $U(1)^N$ gauge transformation $U(\v{k}) = \delta_{ab} e^{i\theta_a(\v{k})}$, Equation \eqref{eq:U(N)_gauge_transformation} reduces to 
\begin{equation}
	\v{A}_{aa} = \v{A}_{aa} - \v{\nabla}_{\v{k}} \theta_{a}(\v{k}).
	\label{eq:abelian_connection_transformation}
\end{equation}
So the Abelian connection transforms as $\v{D} \to \v{D}' = e^{-i\theta_a} \v{D}_{aa} e^{i\theta_a} = \v{D}$ and is thus gauge invariant. This $U(1)^N$ connection is nothing more than one copy of the normal Berry connection for each band. As above, we get an associated connection on operators given by $\v{D}[\widehat{\mathcal{O}}] = [ \v{D}, \widehat{\mathcal{O}}]$, and $\v{D}[\widehat{\mathcal{O}}]$ will be gauge-invariant whenever $\mathcal{O}$ is.

Let us briefly contrast the $U(N)$ and $U(1)^N$ connections and identify when each should be used. The non-Abelian connection $\v{\mathcal{D}}$ is a strictly more complicated object than the Abelian connection $\v{D}$. In general, objects involving $\v{\mathcal{D}}$ will be gauge-covariant after choosing the energy basis, but objects with $\v{D}$ may be gauge-\textit{invariant}. For example, the curvature 
\begin{equation}
	\mathcal{F}^{\v{\mathcal{D}}} = i [\v{\mathcal{D}}, \v{\mathcal{D}}] \to \left(\mathcal{F}^{\v{\mathcal{D}}}\right)' = U^\dagger \mathcal{F}^{\v{\mathcal{D}}} U,
	\label{eq:non-Abelian_curvature}
\end{equation}
is gauge-covariant, whereas in the Abelian case 
\begin{equation}
	F^{\v{D}} = i [\v{D}, \v{D}] \to \left(F^{\v{D}}\right)' = e^{-i\theta_a} F^{\v{D}}_{ab} \delta_{ab} e^{i\theta_b} = F^{\v{D}}
	\label{eq:Abelian_curvature}
\end{equation}	
is gauge-invariant. (This is a standard fact for non-Abelian versus Abelian connections.) As any observable must be strictly gauge-invariant, so it is necessarily much easier to produce observables out of the second connection. In an ideal world, we would be able to work only with $\v{D}$ and not $\v{\mathcal{D}}$. Indeed, for a single band when $N=1$, this is the case. There is some hope of eliminating $\v{\mathcal{D}}$, because for all operators that act diagonally in band space, with $\widehat{\mathcal{O}} = \delta_{ab} \mathcal{O}_{aa}$, the induced connection $\v{\mathcal{D}}$ reduces to $\v{D}$. However, this is a vain hope: measuring electromagnetic responses inevitably involves the off-diagonal components $\v{\mathcal{A}}_{ab}$, and we must use the full generality of the non-Abelian connection. Moreover, when bands are degenerate or cross, such as at a Dirac point, there is no unique way to define the Bloch wavefunctions of each band. These points, which play a crucial role in topological band structures, therefore cannot be fully described via this $U(1)^N$ connection.

In a philosophical sense, the presence of the non-Abelian connection helps to explain why non-linear conductivity responses are often devoid of simple forms: they must be gauge-invariant, but their ``building blocks'' are only gauge-covariant, and so much be composed of tricky combinations that cancel out among themselves. More optimistically, however, one can \textit{harness} this gauge invariance. We will use it to conceptually simplify our perturbation theory approach to non-linear conductivities in the length gauge. A theme from recent years is that the converse is also true: once a new gauge invariant combination has been isolated, it is usually physically measurable, perhaps in a limit. To search for new and interesting quantities to measure, one need only consider what combinations are gauge invariant.

\section{Useful Integrals}
\label{app:integrals}

In this section we will evaluate the loop integrals in the Feynman diagrams. Following Chapter 3 of Mahan \cite{mahan2013many}, we work with Matsubara frequencies, which allows the evaluation of the integrals with straightforward contour integral techniques.

\begin{figure}	
	\centering
	\begin{tikzpicture}
		\draw[latex-latex] (-2.5,0) -- (2.5,0) node[label=right:$\omega$]{};
		\draw[latex-latex] (0,-2.5) -- (0,2.5) node[label=above:$i\omega$] {};
		\foreach \y in {-11,-9,...,11}{
			\node[dot] at (0,0.19*\y) {};
		}
		\node[dot,label=above:{$z_1 = \varepsilon_a$}] at (0.8,0) {};
		\draw (0,0) circle (1.5);
		\draw[-latex] (0,0) -- (225:1.5) node[midway,label=left:$R$] {};
	\end{tikzpicture}
	\caption{Depiction of the poles of the function $f(z) F_1(z)$ and the integration contour. The poles $z_n$ are on the $i\omega$ axis and the pole $z_1$ is on the $\omega$ axis.}
	\label{fig:poles}
\end{figure}
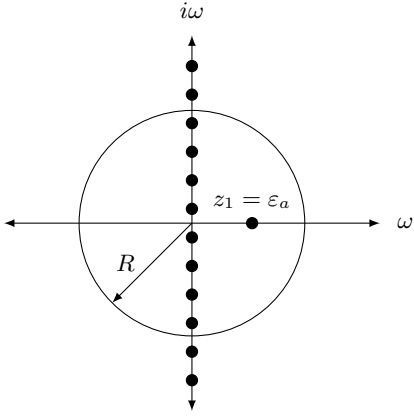

We wish to evaluate integrals such as
\begin{align}
	I_1 \ &=\ \int d\omega\; G_a(\omega) = \int d\omega\; \frac{1}{\omega-\varepsilon_{a}}\\
	I_2(\omega_1) \ &=\ \int d\omega \; G_a(\omega) G_b(\omega+\omega_1)\\
I_3(\omega_1,\omega_2) \ &=\ \int d\omega \; G_a(\omega) G_b(\omega+\omega_1) G_c(\omega + \omega_1 +\omega_2)\\
I_4(\omega_1,\omega_2,\omega_3) \ &=\ \int d\omega \; G_a(\omega) G_b(\omega+\omega_1) G_c(\omega + \omega_1 +\omega_2) \\
\ &\hspace{2em} \times G_d(\omega+\omega_1+\omega_2+\omega_3)
\end{align}
In imaginary time, fermions only have frequencies at odd imaginary integers: $i \omega_n = i(2n+1) \pi/\beta$ for $n \in \Z$. The integral is then analytically continued to a sum
\begin{equation}
	I_1 \to S_1 = \frac{1}{\beta} \sum_{n \in \Z} \frac{1}{i \omega_n - \varepsilon_a}.
\end{equation}
To evaluate this sum, note that the Fermi-Dirac Distribution $f(z) = \frac{1}{e^{\beta z} -1}$ has poles at exactly these complex frequencies $i\omega_n$, each with residue $-1/\beta$. We can therefore use the following trick of trading the sum for a contour integral. Consider
\begin{equation}
	0 = J_1 = \lim_{R\to \infty} \oint_{C_R} \frac{dz}{2\pi i} f(z) F_1(z)
\end{equation}
where the contour is the circle of radius $R$ and
\begin{equation}
	F_1(z) = \frac{1}{z-\varepsilon_a}.
\end{equation}
The integral on the right-hand side is easy to evalute. The poles of $f(z) F_1(z)$, shown in Figure \ref{fig:poles} are at $z_n = i\omega_n$ with residue $R_n = -\frac{1}{\beta} F_1(i \omega_n)$, coming from the Fermi-Dirac distribution, and then $z_1 = \varepsilon_a$ with residue $R_1 = f(\varepsilon_a)$. So
\begin{equation}
	0 = J_1 = -\frac{1}{\beta} \sum_{n \in \Z} F_1(i\omega_n) + f(\varepsilon_a).
\end{equation}
Rearranging,
\begin{equation}
	I_1 = S_1 = f(\varepsilon_a),
\end{equation}
where the first equality is true since the analytic continuation back is trivial here.

Precisely the same technique will work for the more complex integrals with one extra residue for each Green's function. For $I_2$ we analytically continue to 
\begin{equation}
	S_2(i \omega_1) = \frac{1}{\beta} \sum_{n \in \Z} \frac{1}{z-\varepsilon_a} \frac{1}{z + i \omega_1 - \varepsilon_b}.
\end{equation}
Since $i \omega_1$ is due to an incoming photon, it is a bosonic Matsubara frequency and thus an \textit{even} integer instead of odd: $i \omega_1 = i(2M) \pi/\beta$ for some integer $M$. Now consider
\begin{align}
	0 = J_2 &= \lim_{R\to \infty} \oint_{C_R} \frac{dz}{2\pi i} f(z) F_2(z)\\
	F_2(z) &= \frac{1}{z-\varepsilon_a} \frac{1}{z + i \omega_1 - \varepsilon_b}.
\end{align}
The function $f(z) F_2(z)$ has poles and residues

\begin{align}
	z_n &= i \omega_n; &R_n& = -\frac{1}{\beta} F_2(i\omega_n)\\
	z_1 &= \varepsilon_a; &R_1& = \frac{f(\varepsilon_a)}{\varepsilon_a+i\omega_1-\varepsilon_b} = \frac{-f(\varepsilon_a)}{\varepsilon_{ba}-i\omega_1}\\
	z_2 &= \varepsilon_a - i \omega_1; &R_2& = \frac{f(\varepsilon_b - i \omega_1)}{\varepsilon_b-i\omega_1 - \varepsilon_a} = \frac{f(\varepsilon_b)}{\varepsilon_{ba} - i \omega_1}.
\end{align}
In the last equality for $R_2$, the fact $e^{\beta(i\omega_1)} = 1$ implies $f_b(\varepsilon_a-i\omega_1) = f(\varepsilon_b)$. Therefore
\begin{equation}
	S_2(i\omega_1) = R_1 + R_2 = \frac{f_{ab}}{i\omega_1 - \varepsilon_{ab}}.
\end{equation}
Analytically continuing back we then have
\begin{equation}
	I_2(\omega_1) = \frac{f_{ab}}{\omega_1 - \varepsilon_{ab}}.
\end{equation}

The generalization to $I_3$ and $I_4$ follows the same pattern. For $I_3$ we consider the contour integral against
\begin{equation}
	F_3(z) = \frac{1}{z-\varepsilon_a} \frac{1}{z+i\omega_1 - \varepsilon_b} \frac{1}{z+i \omega_{12} - \varepsilon_c}
\end{equation}
where $\omega_{12} = \omega_1 + \omega_2$. Then $f(z) F_3(z)$ has poles and residues
\begin{align}
	z_n &= i \omega_n; &R_n& = -\frac{1}{\beta} F_3(i\omega_n)\\
	z_1 &= \varepsilon_a; &R_1& = \frac{f(\varepsilon_a)}{(\varepsilon_{ab} + i \omega_1)(\varepsilon_{ac} + i \omega_{12})}\\
	z_2 &= \varepsilon_b-i\omega_1; &R_2& = \frac{f(\varepsilon_b)}{(\varepsilon_{ba} - i \omega_1)(\varepsilon_{bc} + i \omega_{2})}\\
	z_3 &= \varepsilon_c-i\omega_{12}; &R_3& = \frac{f(\varepsilon_c)}{(\varepsilon_{ca} - i \omega_{12})(\varepsilon_{cb} - i \omega_{2})}.
\end{align}
Then $S_3(i\omega_1,i\omega_2) = R_1 + R_2 + R_3$. Analytically continuing back to real frequency,
\begin{align}
	\label{eq:I_3}
	&I_3(\omega_1,\omega_2)
	\ =\
	\frac{f(\varepsilon_a)}{(\varepsilon_{ab} + \omega_1)(\varepsilon_{ac} + \omega_{12})}\\
	&- \frac{f(\varepsilon_b)}{(\varepsilon_{ab} +  \omega_1)(\varepsilon_{bc} +  \omega_{2})}
+ \frac{f(\varepsilon_c)}{(\varepsilon_{ac} + \omega_{12})(\varepsilon_{bc} + \omega_{2})}.
\end{align}

Employing the same procedure, $S_4$ is made up of 4 poles, which sum to give

\begin{align}
	\label{eq:I_4}
	&I_4(\omega_1,\omega_2,\omega_3)\\
	\nonumber
	&=\frac{f(\varepsilon_a)}{(\varepsilon_{ab} + \omega_1) (\varepsilon_{ac} + \omega_{12})(\varepsilon_{ad}+\omega_{123})}\\
	\nonumber
	&+\frac{f(\varepsilon_b)}{(\varepsilon_{ba} - \omega_1) (\varepsilon_{bc} + \omega_{2})(\varepsilon_{bd}+\omega_{23})}\\
	\nonumber
	&+\frac{f(\varepsilon_c)}{(\varepsilon_{ca} - \omega_{12}) (\varepsilon_{cb} - \omega_{2})(\varepsilon_{cd}+\omega_{3})}\\
	\nonumber
	&+\frac{f(\varepsilon_d)}{(\varepsilon_{da} - \omega_{123}) (\varepsilon_{db} - \omega_{23})(\varepsilon_{dc}-\omega_{3})}.
\end{align}

\bibliography{references}

\end{document}